\documentclass[a4paper,11pt]{article}
\pdfoutput=1 

\usepackage{jheppub}
\usepackage{xcolor}

\usepackage[T1]{fontenc}
\usepackage{bbold}
\usepackage{esint}
\usepackage{amsmath}
\usepackage{amssymb}
\usepackage{mathtools}
\usepackage{stmaryrd}
\usepackage{tensor}

\usepackage{subcaption}
\usepackage{graphicx}
\usepackage{booktabs}
\usepackage{listings}
\usepackage{inconsolata}
\lstdefinestyle{pycode}{
  language=Python,
  basicstyle=\small\ttfamily,
  keywordstyle=\bfseries\color[rgb]{0.13,0.29,0.53},
  stringstyle=\color[rgb]{0.31,0.60,0.02},
  commentstyle=\color[rgb]{0.56,0.35,0.01},
  emphstyle=\color[rgb]{0.63,0.13,0.13},
  showstringspaces=false,
  breaklines=true,
  frame=single,
  framerule=0.4pt,
  rulecolor=\color{gray!40},
  backgroundcolor=\color{gray!5},
  xleftmargin=1em,
  xrightmargin=1em,
  aboveskip=0.8em,
  belowskip=0.8em,
  lineskip=1pt,
}

\let\<=\langle
\let\>=\rangle

\def\be#1\ee{\begin{align}#1\end{align}}

\newcommand{\al}[1]{\begin{align}#1\end{align}}
\newcommand{\spl}[1]{\begin{split}#1\end{split}}

\newcommand{\tr}{\text{tr}}

\usepackage{tikz}
\usetikzlibrary{calc,decorations.markings}
\tikzset{every picture/.style={line width=0.8pt}}

\tikzset{graph-1/.style = {
  line cap = round,
   line join = round,
     > = triangle 45,
     x=0.7cm, y=0.7cm,
      every node/.append style = {inner ysep=2mm}
                        }
    }
    
\usetikzlibrary{matrix,decorations.pathreplacing}

\title{
On the strong coupling limit of Yang-Mills matrix models
}

\author[a]{Adrien Martina, }
\author[b,c]{Harish Murali }
\affiliation[a]{Fields and Strings Laboratory, Institute of Physics, Ecole Polytechnique Federale de Lausanne (EPFL),
CH-1015 Lausanne, Switzerland}
\affiliation[b]{Perimeter Institute for Theoretical Physics, Waterloo, Ontario N2L 2Y5, Canada}
\affiliation[c]{Department of Physics and Astronomy, University of Waterloo, Waterloo, Ontario, N2L 3G1, Canada}

\abstract{
We study the strong coupling limit of mass deformed Yang--Mills matrix models, with the aim of understanding when the matrices become effectively commuting. The Yang--Mills interaction classically drives the matrices toward mutually commuting valleys, where the matrices can potentially be interpreted as coordinates of an emergent space. However, taking into consideration the integration measure, commutativity is not automatic since the commuting locus is entropically suppressed, and in the bosonic models with $D\geq3$ the strong coupling limit remains non-commuting. We find that fermions change this competition in a sharp way. As the number of fermionic degrees of freedom is increased, there is a critical value $\mathcal N_c=2(D-2)$, realized by the supersymmetric Yang--Mills matrix models, at which the matrices commute at strong coupling. The same critical models also exhibit universality under deformations by $O(N^2)$, or huge, operators: the normalized eigenvalue densities are insensitive to the microscopic details of the huge operators. Increasing the number of fermions beyond the critical point still gives commuting matrices, but the huge-operator universality is lost. Thus commutativity and universality are related but distinct: the matrix models with supersymmetric field content sit at the critical boundary where we have both.
}

\begin{document}
\maketitle
\flushbottom

\newpage
\section{Introduction}
The AdS/CFT correspondence has provided us with a powerful tool to understand quantum gravity by studying the strong coupling limit of certain conformal gauge theories \cite{Maldacena:1997re,Gubser:1998bc,Witten:1998qj,Aharony:1999ti}. However, even for concrete realizations of the duality like $\mathcal N=4$ SYM dual to type IIB on $AdS_5 \times S^5$, it is often challenging to compute generic observables at strong coupling beyond the planar limit. For this reason, there has been a lot of interest in studying zero-dimensional matrix models and matrix quantum mechanics that can capture many essential features of a large $N$ holographic theory. The Ishibashi--Kawai--Kitazawa--Tsuchiya (IKKT) model was proposed as a non-perturbative definition of type IIB string theory \cite{Ishibashi_1997}, and the polarized IKKT model (pIKKT) has recently emerged as a closely related holographic matrix integral with a genuine coupling and well-defined observables \cite{Hartnoll:2024csr,Komatsu:2024bop,Komatsu:2024ydh,Hartnoll:2025ecj}. Their simplicity and tractability using both numerical and analytical methods makes them excellent laboratories for studying the emergence of spacetime and the dynamics of quantum gravity in a holographic setting.

\begin{figure}
    \centering
    \includegraphics[width=0.8\linewidth, trim={0cm, 6.5cm, 0cm, 0cm}, clip]{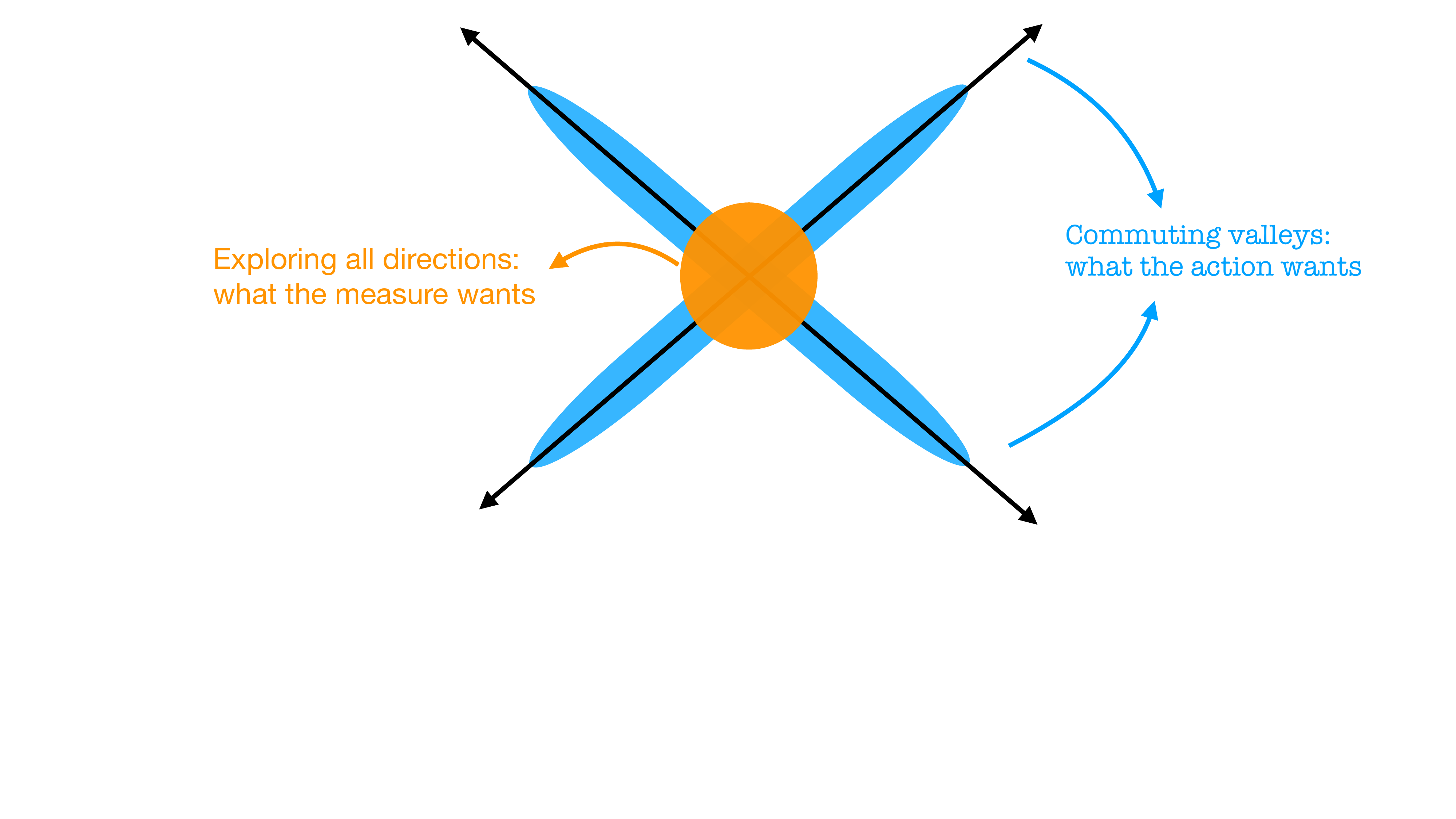}
    \caption{A schematic depiction of the configuration space of the matrices. There are two competing effects: the Yang-Mills term $-\tr [X_I, X_J]^2$ in the action pushes the matrices towards the commuting directions. On the other hand such configurations are very non-generic with the off-diagonals close to zero and therefore their measure is small.}
    \label{fig:measureVsAction}
\end{figure}

The question we focus on in this paper is whether the matrices become commuting in the holographic strong-coupling regime (i.e. the {}'t~Hooft limit, with large {}'t~Hooft coupling), and how robust this phenomenon is for a range of models. Both IKKT and pIKKT are Yang-Mills type matrix models, in the sense that their bosonic action contains a commutator-squared interaction of the form
\begin{equation}
    \frac{N}{\lambda}\,
    \operatorname{tr}\left(-\frac14\sum_{I,J}[X_I,X_J]^2\right),
    \label{eq:intro-ym-term}
\end{equation}
where $X_I$ are Hermitian matrices. Since $[X_I,X_J]$ is anti-Hermitian, the term in \eqref{eq:intro-ym-term} is non-negative and is minimized by mutually commuting matrices. If the matrices commute, they can be simultaneously diagonalized and their eigenvalues can potentially admit an interpretation as coordinates of an emergent space. However, commuting configurations occupy a small region of the full matrix configuration space, and the entropic suppression may overcome the attraction towards the commuting locus, see figure \ref{fig:measureVsAction}. The main conclusion from our numerical experiments is that this competition is strongly influenced by the number of fermions: too few fermions leave the strong-coupling theory non-commuting, while the theory becomes commuting with sufficient number of fermions. The critical point of this phase transition is when we have $\mathcal N_c = 2(D-2)$ real fermionic degrees of freedom for $D$ bosonic degrees of freedom. This is realized in $D=3,4,6$ and $10$ by the supersymmetric Yang-Mills models. Supersymmetry itself is not the essential ingredient; its role in these examples is to place the theory at the critical fermion number. We check this directly by explicitly breaking supersymmetry in the 4D Type~I model, while keeping the same fermionic content, and find the same commuting strong-coupling behavior.

A simple example of the non-triviality of competition between classical commuting valleys and entropy is provided by the mass deformed bosonic Yang-Mills matrix models, whose partition function reads
\begin{equation}
    Z = \int \prod_{I=1}^D dX_I \exp\left[-\frac N\lambda\,\tr \left( \sum_I X_I^2 - \frac{1}{4} \sum_{IJ}[X_I,X_J]^2\right)\right], \label{eq:bosonicYMAction}
\end{equation}
where $X_I$ are $D$ traceless hermitian matrices of size $N\times N$. Note here that the mass deformation is crucial in order to have a sensible notion of strong coupling. Without the $\tr X^2$ term, we can rescale the matrices and get rid of the coupling. The $D=2$ model, also called the Hoppe model, can be solved by integrating out $X_2$ \cite{Kazakov:1998ji,hoppe:1989,Berenstein:2008eg}. On the other hand, the $D\geq 3$ model is not analytically solvable. However, it is quite straightforward to simulate it numerically using Monte Carlo -- see Appendix \ref{app:hmc} for a brief review. A convenient diagnostic of commutativity is the ratio of the commutator sqaured to the anticommutator squared, which is invariant under rescaling of the matrices: 
\begin{equation}
    R = \left \langle\frac{-\sum_{I<J} \tr [X_I,X_J]^2}{\sum_{I<J} \tr \{X_I,X_J\}^2}\right \rangle\label{eq:commAnticommRatio}
\end{equation}

\begin{figure}
    \centering
    \includegraphics[width=0.6\textwidth]{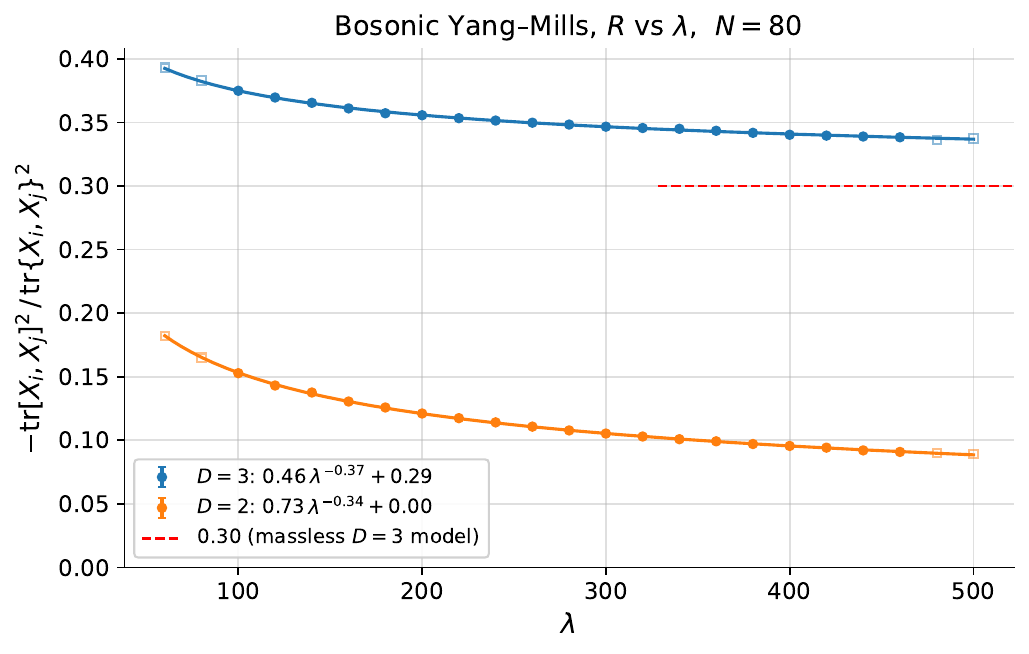} 
    \caption{Monte Carlo results for the commutativity ratio $R$ for the 2D and 3D bosonic Yang-Mills models at $N=80$, over the range $\lambda\in[20,500]$. The data is shown along with power law fits of the form $a + b\lambda^{-c}$. Open squares mark points that were not used for the fit, and show that the fits are reliable. In the $D=2$ case, it decays to zero as a power law $R\sim\lambda^{-0.34}$ whereas in $D=3$, $R$ saturates to an $\mathcal{O}(1)$ constant $\approx 0.30$. The asymptote is close to the value of $R$ in the massless bosonic $D=3$ model at $N=80$.}
    \label{fig:HoppeVs3D}
\end{figure}

Figure \ref{fig:HoppeVs3D} shows the plot of $R$ as a function of the coupling $\lambda$ for $D=2$ and $D=3$ at $N=80$. The ratio $R$ behaves very differently for these two cases. For $D=2$, it decays as a pure power law $R\sim\lambda^{-\frac13}$. It becomes arbitrarily small at strong coupling, indicating that this model is commuting. We review the analytic derivation of the exponent $\lambda^{-\frac13}$ in Appendix \ref{app:hoppe}. On the other hand, for $D=3$, the ratio asymptotes to a constant $O(1)$ non-commutativity at strong coupling. The same behavior persists for other $D>3$. In the competition between entropy and commuting valleys, entropy wins for the bosonic Yang-Mills model in $D\geq3$, but not in $D=2$. A closely related difference between these two cases is that for $D\geq3$, the integral is finite and well defined even without the mass term, whereas for $D=2$, the massless model is divergent. As this example illustrates, commutativity at strong-coupling in a large-$N$ theory is a dynamical and model-dependent phenomenon.

What sort of terms can we add to the action \eqref{eq:bosonicYMAction} to make it commuting for $D\geq3$? We need to enhance the repulsion between the eigenvalues of the matrices. One way to achieve this is to introduce fermions. To see why, consider the following toy model with one bosonic and one Grassmann matrix
\begin{equation}
    S = N\,\tr \left( \frac12 X^2 + \psi^\dagger \psi +i\lambda \psi^\dagger[X,\psi]\right).
\end{equation}
Integrating out $\psi$, we get a Vandermonde-like factor $\prod_{i<j}(1+\lambda^2 (x_i - x_j)^2)$, which strengthens the eigenvalue repulsion. In Yang--Mills matrix models, this fermion-induced repulsion can help overcome the entropic suppression of commuting configurations and drive the matrices toward the commuting valleys. This suggests that supersymmetric Yang--Mills matrix models are good candidates for commutativity.

In section \ref{sec:susyModels} we therefore study the mass deformed supersymmetric Yang--Mills matrix models classified in \cite{Martina:2025kwc}. In particular, we look at the $D=3$ and $D=4$ models which have fewer fermions and thus are more tractable numerically. We restrict our attention to models without a Myers term, and thus without fuzzy sphere saddles\footnote{The stability of the classical fuzzy sphere saddles in the strong coupling regime will be analyzed in an upcoming work \cite{ourUpcomingPaper}. Unlike in the BMN model where fuzzy spheres are BPS vacua \cite{Dasgupta:2002ru,Dasgupta:2002hx,Kim:2002if}, their stability in matrix models is not guaranteed.}. Within this class, the 4D Type~I model is special because the Pfaffian obtained by integrating out the fermions is positive definite. The $D=3$ model on the other hand does have a sign problem, and we study the phase quenched model.

We also study universality under deformation of the integrals by heavy insertions in section \ref{sec:universality}. In holography, operators whose charges or dimensions scale as $O(N^2)$ backreact on the dual geometry and are expected to be described, at least in typical regimes, by a small number of coarse-grained parameters. A matrix-model version of this idea was recently studied in \cite{Guerrieri:2025ytx}, where it was found that the eigenvalue distribution and probe correlators in the presence of such huge operators in the Hoppe model were universal and were completely determined by a few moduli at strong coupling -- analogous to the black hole no-hair theorem. One of the motivations of this work was to see if these nice features of commutativity and universality are unique to the Hoppe model or if they extend to other multi-matrix models. We find that the 4D Type~I and the 3D SUSY models also exhibit this universality under deformations by different $O(N^2)$ sources.

In section \ref{sec:numFermions}, we study what happens as we change the number of fermionic degrees of freedom. To have a notion of continuously varying the fermion number, we simply raise the fermionic determinant to an arbitrary power and analytically continue in this variable. We find that when there are too few fermions, like in the $D\geq3$ bosonic models, the corresponding massless models are finite and the strong coupling limits asymptote to these models. Since the massless models have no free parameters, the strong coupling limit is not commuting in this case. As we increase the number of fermions, there exists a critical point with $\mathcal N_c=2(D-2)$ fermionic degrees of freedom where the massless integral is barely convergent, or even divergent as in $D=2$ and phase quenched $D=3$ \cite{Austing:2001bd, Austing:2001ib, Krauth_1998, Krauth:1999qw}. In $D=3,4,6$ and $10$, this is the supersymmetric point, and the eigenvalue distributions in these models have power-law tails \cite{Krauth:1999qw}. We find that these models with $2(D-2)$ fermions when mass deformed lead to commuting matrices at strong coupling.

Increasing the fermion number beyond $\mathcal N_c$ also leads to commuting matrices, and in this regime we have a simple analytic description. At large $\lambda$ the off-diagonal modes become heavy around a configuration of well-separated eigenvalues, and can be integrated out perturbatively. This gives an effective theory for the $N$ diagonal eigenvalue vectors. The fermionic modes and gauge fixing terms combine into a generalized Vandermonde factor in the effective measure, with power $\mathcal{N}-\mathcal{N}_c$. Thus for $\mathcal{N}>\mathcal{N}_c$ the diagonal theory contains explicit eigenvalue repulsion. This repulsion keeps the off-diagonal masses large and the diagonal approximation is self-consistent even in the 't~Hooft limit. At the critical point, by contrast, the Vandermonde power vanishes. The diagonal approximation is then only self-consistent at strong coupling for finite $N$. In the supercritical effective theory one can compute observables analytically, in good agreement with our numerics. Since the coupling can once again be absorbed by rescaling the variables in this effective theory, a huge $O(N^2)$ deformation changes the eigenvalue distribution in a non-universal way. 

All the numerical results in this work were obtained with \href{https://github.com/harish02murali/matrix-hmc}{\texttt{matrix-hmc}}, a GPU-accelerated Hybrid Monte Carlo (HMC) package developed by the authors using PyTorch. The package is designed to be easy to use, efficient, and can be extended to general multi-matrix models. We make it publicly available\footnote{\url{https://github.com/harish02murali/matrix-hmc}} as a reusable tool for numerical studies of matrix models.

\section{Supersymmetric models}
\label{sec:susyModels}
In the absence of any mass-deformations, the SUSY Yang-Mills models take the form
\begin{equation}
    S_{\text{massless}} = \frac{N}{\lambda} \tr\left(-\frac14 [X_I,X_J]^2 -\frac i2 \bar\psi\Gamma^I[X_I,\psi]\right)\label{masslessSUSY}
\end{equation}
where $X_{I=1\ldots D}$ are $D$ traceless hermitian matrices and $\psi_{\alpha=1\ldots\mathcal N}$ are $\mathcal N=2(D-2)$ hermitian fermionic matrices. In this work, we will follow the Gamma matrix basis and conventions from \cite{Martina:2025kwc}. The SUSY Yang-Mills models are defined in dimensions $D=3,4,6$ and $10$ and have $\mathcal N$ supersymmetries \cite{Krauth:1998xh}. The $D=10$ case is the Ishibashi-Kawai-Kitazawa-Tsuchiya (IKKT) model \cite{Ishibashi_1997} which is conjectured to be dual to Type IIB string theory in 10D. The lower $D$ versions are also interesting toy models to study and constitute a great playground to develop our intuition. Note that there are no free parameters in \eqref{masslessSUSY} -- the coupling $\lambda$ can be absorbed by rescaling the matrices. So, there is no notion of a strong coupling limit.

In \cite{Martina:2025kwc}, deformations of \eqref{masslessSUSY} were classified and those models have genuine couplings that cannot be rescaled away. In particular, the $D=4$ models and their mass deformations are very interesting, since they are free from the fermion sign problem \cite{Krauth:1998xh, Ambjorn:2000bf, Martina:2025kwc}. In what follows, we study the strong coupling limit in the four- and three-dimensional massive SYM models.

\subsection{4D type I model}
In four dimensions, there are two different mass deformations that preserve $\mathcal N=4$ supersymmetry, referred to as Type I and Type II in \cite{Martina:2025kwc}. Here, we focus on the Type I model, which has the action
\begin{equation}
    S_{D=4,\text{type I}} =\frac{N}{\lambda}\text{tr }\Biggl(-\frac14 [X_I,X_J]^2 -\frac{i}{2}\bar\psi \Gamma^I [X_I,\psi] + X_I^2 + \bar\psi\psi \Biggr) \label{eq:typeIaction}
\end{equation}
We can integrate out the fermions and obtain an effective theory for the bosonic degrees of freedom
\begin{equation}
    Z_{\text{eff}} = \int dX_I\ \text{Pf}[\mathcal M(X)]\exp\left[\frac{N}{\lambda}\text{tr}\Bigl(\frac14 [X_I,X_J]^2 - X_I^2 \Bigr)\right]
\end{equation}
where $\text{Pf}[\mathcal M(X)]$ is the Pfaffian of a $4(N^2-1)\times 4(N^2-1)$ matrix obtained from integrating out real fermions. Up to a change of basis, the matrix is given by
\begin{equation}
    \mathcal M(X) = \frac12\begin{pmatrix}
0 & 2 i  & -\mathbf{X}_3 - i \mathbf{X}_4 & -\mathbf{X}_1 + i \mathbf{X}_2 \\
-2 i  & 0 & -\mathbf{X}_1 - i \mathbf{X}_2 & \mathbf{X}_3 - i \mathbf{X}_4 \\
-\mathbf{X}_3 - i \mathbf{X}_4 & -\mathbf{X}_1 - i \mathbf{X}_2 & 0 & -2 i \\
-\mathbf{X}_1 + i \mathbf{X}_2 & \mathbf{X}_3 - i \mathbf{X}_4 & 2 i & 0
\end{pmatrix}.
\end{equation}
where $(\mathbf{X}_I)_{AB} = f_{ABC}\,X_I^C$ are $(N^2-1)\times(N^2-1)$ matrices in the adjoint, $X_I = X_I^CT_C$ and $f_{ABC}$ are the structure constants, $[T_A,T_B]=if_{ABC}T_C$. This Pfaffian is positive definite and can be written as a manifestly positive determinant of half the size \cite{Martina:2025kwc} (this makes a huge difference for the numerics)
\begin{equation}
    \begin{aligned}
    \text{Pf}[\mathcal M(X)] &= \sqrt{\det(\mathbb{1}_{2(N^2-1)} + H^\dagger H)},\\
    H(X) &= \frac12 \begin{pmatrix}
\mathbf{X}_1 - i \mathbf{X}_2 & -\mathbf{X}_3 - i \mathbf{X}_4 \\
-\mathbf{X}_3 + i \mathbf{X}_4 & -\mathbf{X}_1 - i \mathbf{X}_2
\end{pmatrix}.
    \end{aligned}\label{eq:fermionDetTypeI}
\end{equation}

We can do an honest Monte Carlo simulation of this matrix integral with positive measure. The most computationally expensive part of HMC is the evaluation of the $O(N^2)\times O(N^2)$ determinant \eqref{eq:fermionDetTypeI} and its derivative with respect to $X_I$. These steps scale as $O(N^6)$, making it prohibitively expensive for large $N$.\footnote{One can use the Rational Hybrid Monte Carlo method here which is much cheaper in terms of resources, but with more moving parts and parameters to tune. We stick to the vanilla HMC in this work.} Note that in many simpler bosonic models, including the Yang-Mills models studied at the beginning of this section, we only deal with $N\times N$ matrices and matrix operations on those, like multiplication or eigen-decomposition, which scale as $O(N^3)$. For the SUSY models, HMC at $N\sim 50$ was only feasible with modern, highly capable GPUs developed for AI workloads.

\begin{figure}[t]
    \centering
    \begin{subfigure}[c]{0.44\textwidth}
        \centering
        \includegraphics[width=\textwidth]{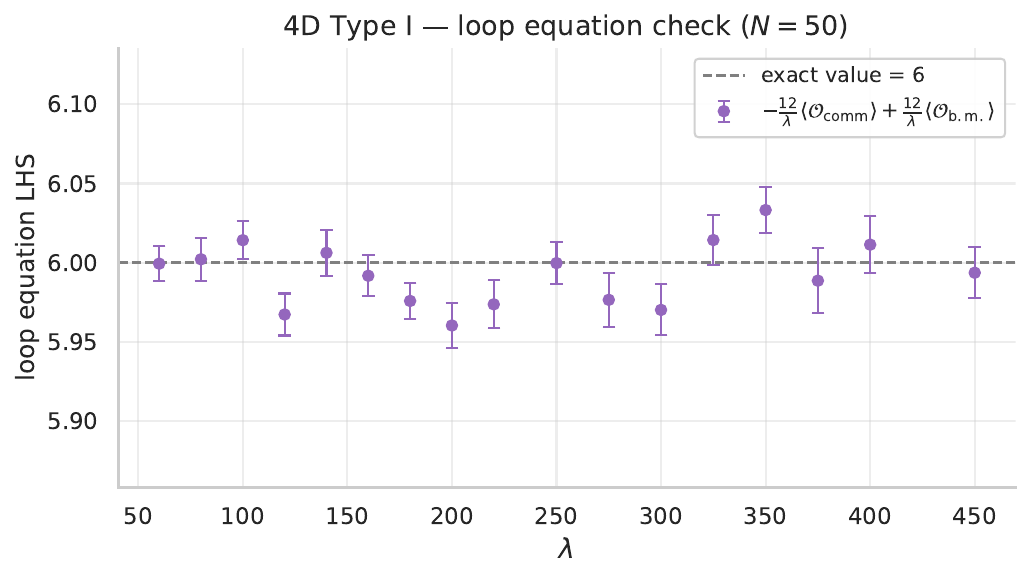}
    \end{subfigure}
    \hfill
    \begin{subfigure}[c]{0.54\textwidth}
        \centering
        \includegraphics[width=\textwidth]{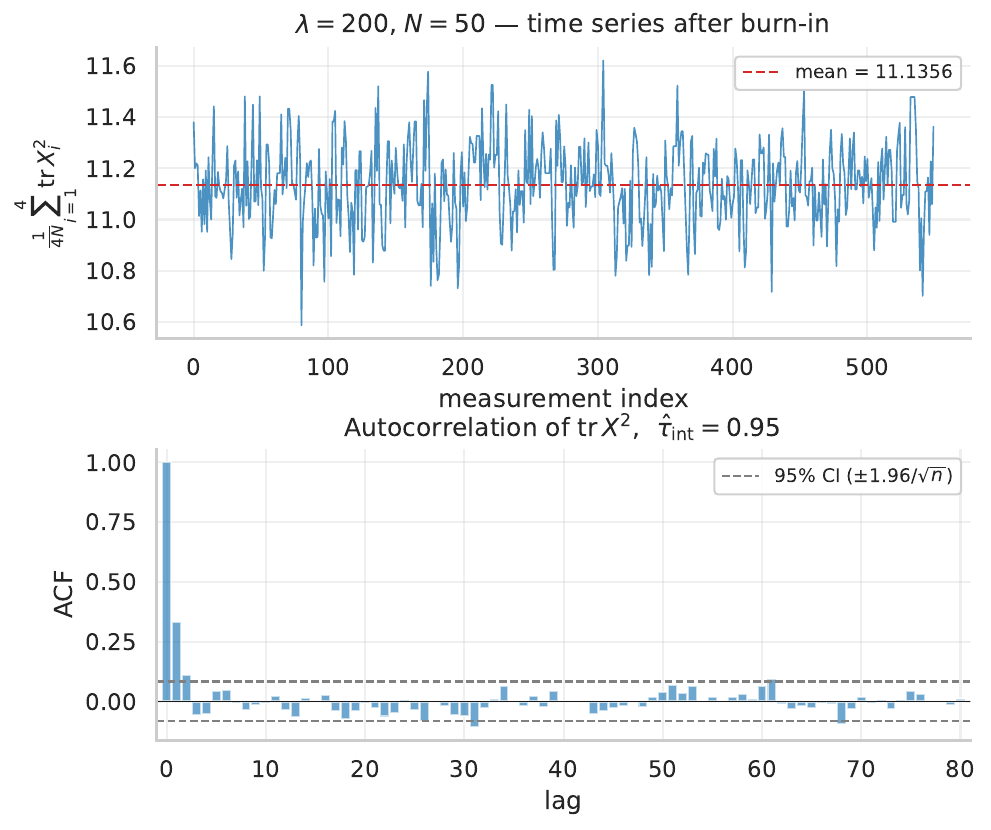}
    \end{subfigure}
    \caption{HMC validation for the 4D Type~I model at $N=50$. \textbf{Left:} The bosonic loop equation \eqref{eq:bosonicType1Constraint} is satisfied across $\lambda\in[60,450]$. \textbf{Right:} At $\lambda=200$, the time series of $\frac{1}{4N}\sum_i\tr X_i^2$ shows no drift and the autocorrelation drops to zero within a few HMC steps.}
    \label{fig:type1-validation}
\end{figure}

Since this is the first non-trivial model we study in this paper, let us be a bit more careful and verify that the HMC is sampling the correct distribution. Two simple checks are shown in Figure~\ref{fig:type1-validation}. One can derive a bosonic loop equation for the Type~I action
\begin{equation}
    -\frac{1}{N\lambda} \langle \tr [X_I, X_J]^2\rangle + \frac{3}{N\lambda} \langle \tr X_I^2 \rangle = 6\left(1-\frac1{N^2}\right),\label{eq:bosonicType1Constraint}
\end{equation}
obtained by taking a linear combination of bosonic and fermionic variations. It is satisfied within statistical errors across the full scan, confirming the normalization and force implementation. Second, at a representative coupling $\lambda=200$, the time series of $\frac{1}{4N}\sum_i \tr X_i^2$ shows no visible drift. We also compute the normalized autocorrelation function
\begin{equation}
    \mathrm{ACF}(\tau)=
    \frac{\langle \delta O_t\,\delta O_{t+\tau}\rangle}
    {\langle \delta O_t^2\rangle}\,,\qquad
    \delta O_t=O_t-\langle O\rangle ,
\end{equation}
which measures how strongly two measurements separated by $\tau$ HMC steps are correlated. The ACF drops to zero within a few steps, so successive measurements are effectively independent.

\begin{figure}[t]
    \centering
    \begin{subfigure}[t]{0.56\textwidth}
        \centering
        \includegraphics[width=\textwidth]{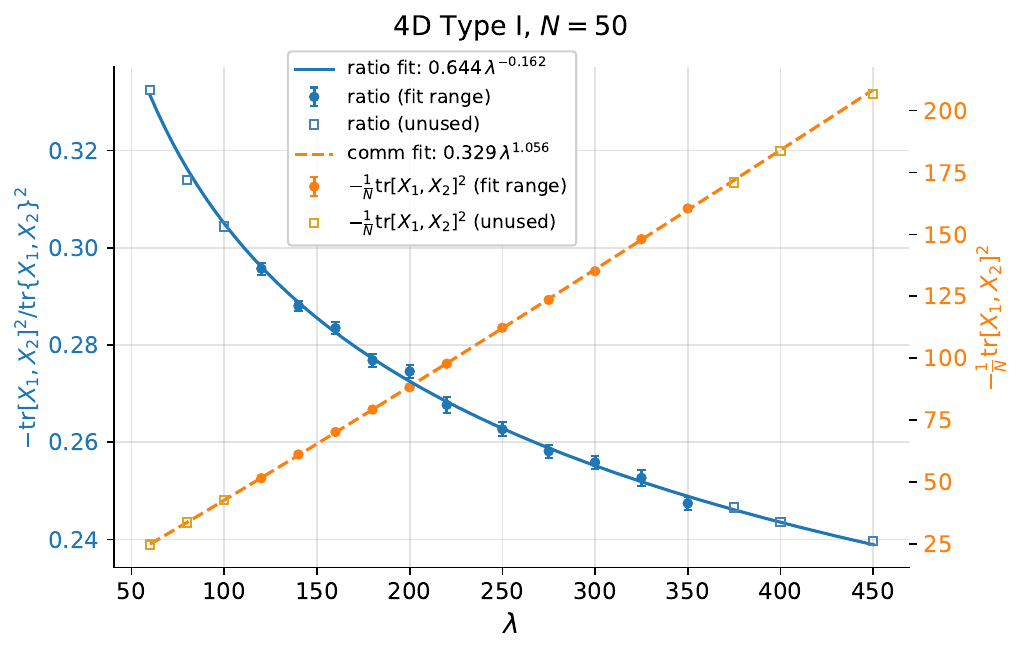}
    \end{subfigure}
    \hfill
    \begin{subfigure}[t]{0.41\textwidth}
        \centering
        \includegraphics[width=\textwidth]{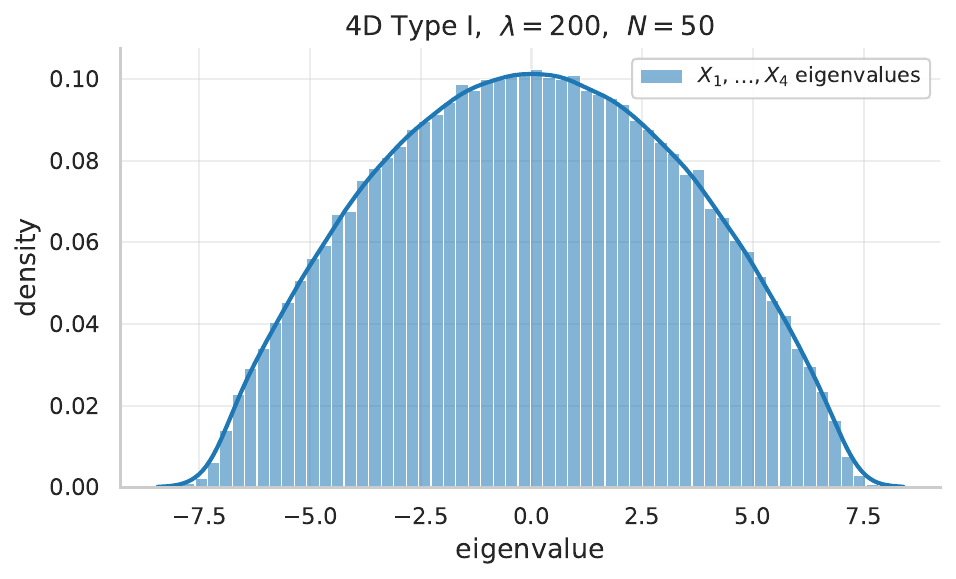}
    \end{subfigure}
    \caption{Results for the 4D Type~I model at $N=50$. \textbf{Left:} The ratio of commutator to anti-commutator (left axis, blue) and commutator magnitude (right axis, orange) vs.\ $\lambda$. Filled circles: points used in the fit. Open squares: withheld cross-validation points. Power-law fits give exponents $-0.16$ and $+1.06$ respectively. \textbf{Right:} Eigenvalue distribution of $X_1,\ldots,X_4$ at $\lambda=200$, $N=50$. The density is smooth and finitely supported. Furthermore, the shape of the density is coupling independent at strong coupling and only the overall scale changes with $\lambda$.}
    \label{fig:type1-results}
\end{figure}

With the simulation validated, we turn to the physics. The HMC scan over $\lambda\in[60,450]$ at $N=50$ is summarised in Figure~\ref{fig:type1-results}. The left panel shows the commutativity ratio $R$ \eqref{eq:commAnticommRatio} and the commutator magnitude $-\frac{1}{N}\tr[X_I,X_J]^2$ as functions of $\lambda$, together with power-law fits. To assess robustness, the first and last three points are withheld from the fit and serve as a cross-validation set; the fitted curves extrapolate to those points nicely. The ratio decays as $R\sim\lambda^{-0.16}$ with no sign of saturation, providing clear evidence that the 4D Type~I model enters a commuting phase at strong coupling. The commutator magnitude itself grows linearly with $\lambda$, keeping the Yang-Mills term in the action $\frac{1}{N} \frac{1}{\lambda} \langle\mathrm{tr} [X_I,X_J]^2 \rangle \sim O(1)$ in $\lambda$. In Appendix~\ref{app:type1Fits} we compare a battery of functional forms; the pure power law gives the best $\chi^2/\mathrm{dof}$. The right panel is the distribution of $X_1,\ldots,X_4$ eigenvalues at $\lambda=200$, which shows that the eigenvalue density is smooth and finitely supported.

\subsection{3D SUSY model}
\label{subsec:3dSUSY}
In \cite{Martina:2025kwc}, the following two parameter mass deformation of the 3D Yang-Mills model which preserves $\mathcal N=2$ SUSY was found
\begin{equation*}
    S_{D=3} =N\ \text{tr }\Biggl(-\frac14 [X_I,X_J]^2 -\frac{i}{2}\bar\psi \Gamma^I [X_I,\psi] + \mu_1\mu_2X_I^2+i\frac{2}3(\mu_1+\mu_2)\epsilon_{IJK}X_I X_JX_K + i\mu_1\bar\psi\psi \Biggr)
\end{equation*}
\begin{figure}
    \centering
    \begin{subfigure}[c]{0.55\textwidth}
        \centering
        \includegraphics[width=\textwidth]{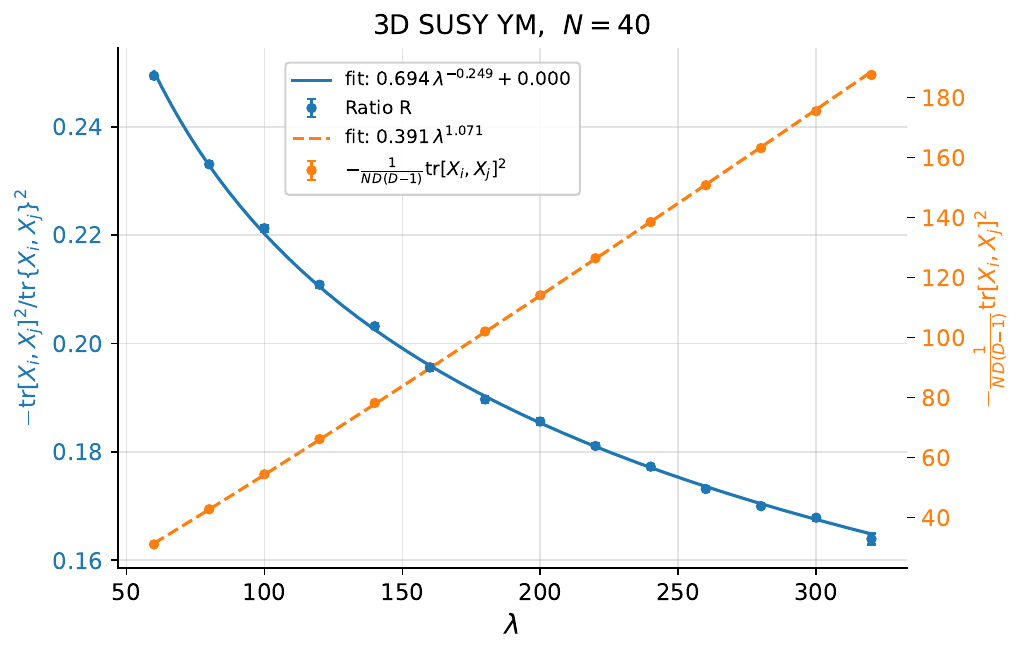}
    \end{subfigure}
    \hfill
    \begin{subfigure}[c]{0.44\textwidth}
        \centering
        \includegraphics[width=\textwidth]{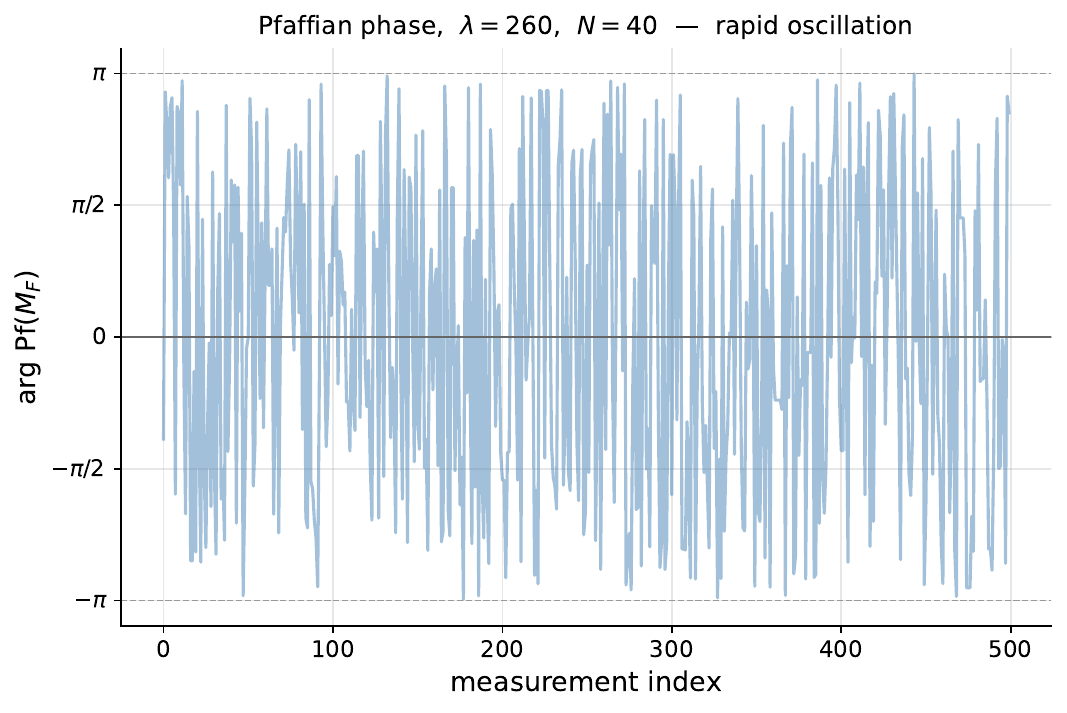}
    \end{subfigure}
    \caption{Phase-quenched $D=3$ massive SYM model at $N=40$. \textbf{Left:} Commutativity ratio $R$ \eqref{eq:commAnticommRatio} and commutator magnitude versus $\lambda$; the fit gives $R\simeq0.694\,\lambda^{-0.249}$. \textbf{Right:} Phase of the unquenched Pfaffian at $\lambda=260$, showing rapid fluctuations across $[-\pi,\pi]$.}
    \label{fig:3dSUSY}
\end{figure}

Here, we focus on the case $\mu_2=-\mu_1=i\Omega$, which kills the Myers term and the fuzzy sphere vacua. After rescaling the matrices $(\lambda \equiv N/ \Omega^4)$ and integrating out the fermions, we end up with the integral
\begin{equation}
    Z_{D=3} = \int dX_I\ \text{Pf}[\mathcal M_{3D}(X)] \exp\left[\frac{N}{\lambda}\text{tr}\Bigl(\frac14 [X_I,X_J]^2 - X_I^2 \Bigr)\right]
\end{equation}
where $\mathcal M_{3D}$ is a $2(N^2-1)\times2(N^2-1)$ matrix given by
\begin{equation}
    \mathcal M_{3D} = \frac{-i}2 \begin{pmatrix}
\mathbf{X}_1 + i \mathbf{X}_3 & 2-\mathbf{X}_2 \\
-2-\mathbf{X}_2 & -\mathbf{X}_1 + i \mathbf{X}_3
\end{pmatrix}
\end{equation}
Unlike in the $4D$ SUSY models, the above Pfaffian is not real for Hermitian $X_I$. We can only simulate the phase quenched model in our numerics where we replace 
\begin{equation}
    \text{Pf}[\mathcal M_{3D}(X)] \rightarrow \biggl|\text{Pf}[\mathcal M_{3D}(X)]\biggr|
\end{equation}
We ran HMC for this phase quenched model for various couplings $\lambda\in[60, 300]$ and $N=40$. As seen in figure \ref{fig:3dSUSY}, the ratio of commutator to anticommutator \eqref{eq:commAnticommRatio} follows a power law $R_{3D}\sim \lambda^{-0.25}$, demonstrating that this model is also commuting at strong coupling. In figure \ref{fig:3dSUSY}, we also plot the phase of the Pfaffian at each Monte Carlo step for $\lambda=260$. The phase is wildly fluctuating, which shows that the phase quenched model is very different from the unquenched case. 

\begin{figure}
    \centering
    \includegraphics[width=0.6\textwidth]{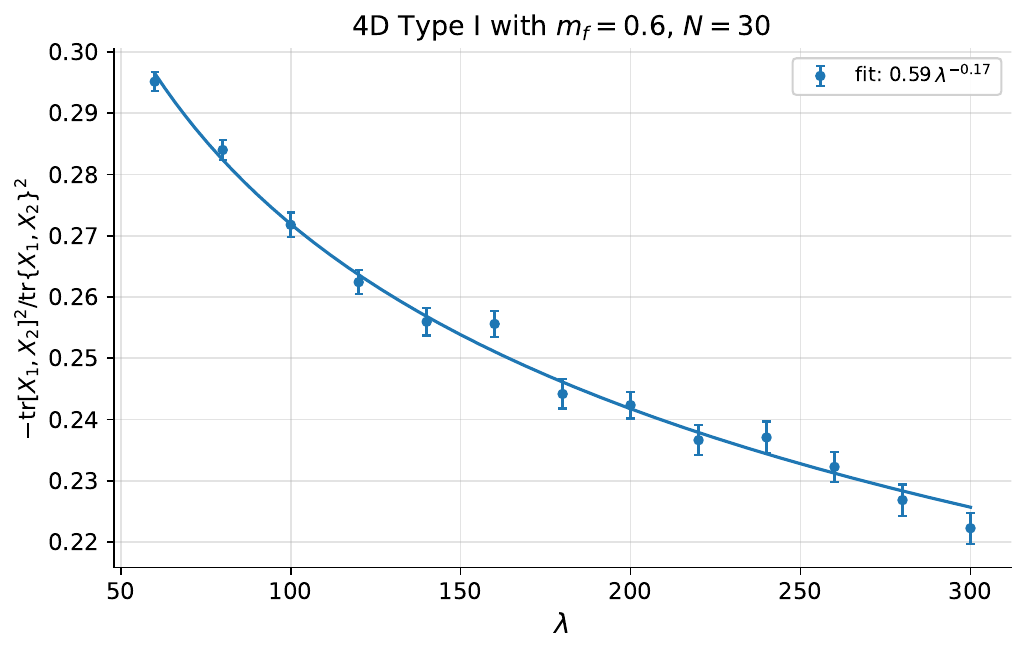}
    \caption{The ratio $\left(-\frac{\tr[X_1,X_2]^2}{\tr\{X_1,X_2\}^2}\right)$ vs.\ $\lambda$ for the 4D Type~I model at $N=30$ with a modified fermion mass term $m_f=0.6$. The power-law fit gives an exponent of $-0.17$, close to the SUSY case $m_f=1$.}
    \label{fig:typeI_fm0.6}
\end{figure}

We have seen that mass deformed models with supersymmetry in three (phase quenched) and four dimensions are commuting at strong coupling. On the other hand, their bosonic cousins are not commuting. So, a natural question that arises is, how important is SUSY in order to have a commuting strong coupling limit? To test this, we can deform the $4D$ Type~I model by changing the fermion mass term, $\tr\ \bar\psi\psi \rightarrow m_f\ \tr\ \bar\psi\psi$. For $m_f\neq1$ the SUSY is explicitly broken. The Monte Carlo results with this deformation are shown in figure \ref{fig:typeI_fm0.6} which shows that the ratio of commutator and anti-commutator scales as $\lambda^{-0.17}$ -- commutativity is preserved even with the deformation. Also, note that in the case of the 3D SUSY model, we have a severe sign problem and we studied the phase-quenched model. However, the model is only supersymmetric with the inclusion of the phase of the Pfaffian. This suggests that the relevant ingredient for commutativity is the presence of sufficient fermionic repulsion, rather than supersymmetry itself.

\subsection{Joint densities}

We now explain why the vanishing of the commutativity ratio \(R\) implies a stronger notion of effective commutativity. Namely, we would like single-trace words in the matrices to become independent of their ordering at strong 't Hooft coupling. We show this under the assumption, satisfied in the massive Yang--Mills matrix models considered here, that the rescaled single-matrix eigenvalue density has finite moments.

Let \(R\sim \lambda^{-\alpha}\), and assume that the one-matrix eigenvalue density has support of size \(L\sim \lambda^\beta\). Using
\[
    -\frac1N\langle \tr [X_I,X_J]^2\rangle\sim \lambda,
    \qquad
    \frac1N\langle \tr \{X_I,X_J\}^2\rangle\sim L^4,
\]
one obtains \(\beta=(1+\alpha)/4\). We therefore introduce
\[
    X_I=\lambda^{(1+\alpha)/4}Y_I,
\]
so that \(\frac1N\langle\tr Y_I^2\rangle=O(1)\) and
\[
    -\frac1N\langle\tr [Y_I,Y_J]^2\rangle\sim \lambda^{-\alpha}.
\]
For any word \(Z\) in the \(Y_I\)'s, using
\(\frac{1}{N}\langle \tr Z^\dagger Z\rangle\sim O(1)\), the Cauchy--Schwarz inequality gives
\[
    \left|
    \left\langle \frac1N\tr [Y_1,Y_2]Z\right\rangle
    \right|
    \leq
    \left[
    \left\langle\frac1N\tr Z^\dagger Z\right\rangle
    \left\langle-\frac1N\tr [Y_1,Y_2]^2\right\rangle
    \right]^{1/2}
    =O(\lambda^{-\alpha/2}).
\]
Thus adjacent matrices in any fixed single-trace word can be exchanged up to corrections of order \(O(\lambda^{-\alpha/2})\). The leading strong-coupling single-trace correlators are therefore commutative.

Like in the Hoppe model \cite{Berenstein:2008eg}, we can then define a joint density that computes the mixed moments
\begin{equation}
    \frac1N\langle\tr\, X_1^{n_1}X_2^{n_2}X_3^{n_3} \dots X_D^{n_D}\rangle = \int \prod_{i=1}^D dx_i\, \rho_D(x_1,\ldots,x_D)\, x_1^{n_1}\cdots x_D^{n_D},\label{eq:rho4Defn}
\end{equation}
at leading order in the 't Hooft limit. The density $\rho_D(\vec x)$ is the unique $SO(D)$ invariant uplift of the eigenvalue density $\rho(x)$ of a single matrix. It is only a function of $r= |\vec x|$, $\rho_D(\vec x) = \rho_D(r)$ and satisfies the marginalization property
\begin{align}
\begin{split}
    \rho(x_1)
=
\int d^{D-1}y\,
\rho_D\!\left(\sqrt{x_1^2+|\vec y|^2}\right)
=
\Omega_{D-2}
\int_{|x_1|}^{\infty}
dr\, r\,(r^2-x_1^2)^{(D-3)/2}\rho_D(r).
\end{split}
\end{align}
where $\Omega_{D-2}$ is the volume of the unit $(D-2)$-sphere.

\begin{figure}
    \centering
    \includegraphics[width=0.8\linewidth]{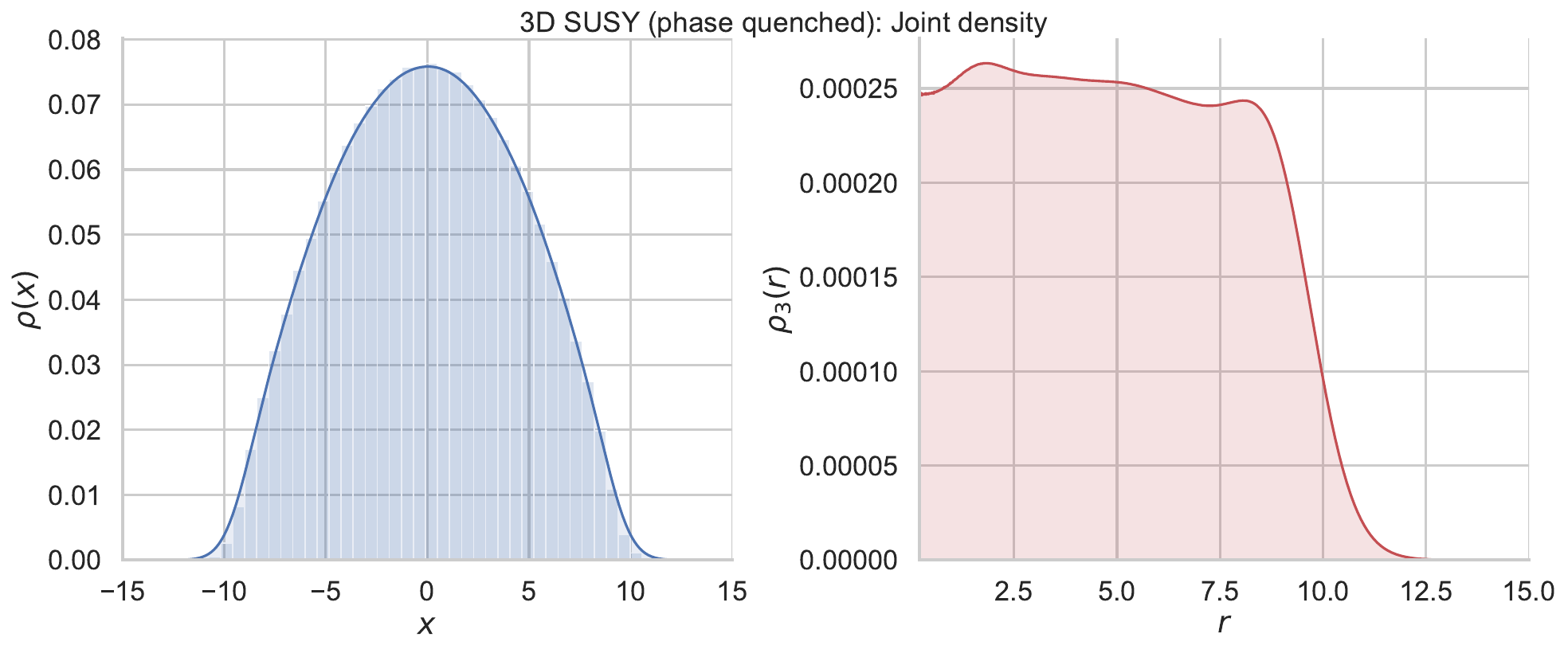}
    \caption{On the left, we have the density of eigenvalues of a single matrix in the phase quenched 3D SUSY model at $\lambda=300$ and $N=40$. The solid line is a smoothed Gaussian KDE, using which we estimate $\rho_3(r) = -\frac{\rho'(r)}{2\pi r}$ shown on the right. Note that the eigenvalue density $\rho(x)$ looks approximately like a parabola for which the rotational invariant $\rho_3$ would be a step function. As seen from the figure on the right, the joint density resembles a smoothed step function.}
    \label{fig:3dJointDensity}
\end{figure}

The function that satisfies this property is obtained by an inverse Abel-type transform,
\begin{equation}
    \rho_D(r) = \begin{cases}
        \left( - \frac{1}{2 \pi r} \frac{d}{dr} \right)^m \rho(r) & D = 2m + 1 \quad m \in \mathbb{N} \\
        \left( - \frac{1}{2 \pi r} \frac{d}{dr} \right)^{m-1} \left[- \frac{1}{\pi} \int_r^\infty dx \frac{\rho'(x)}{\sqrt{x^2-r^2}} \right] & D=2m \quad m \in \mathbb N
    \end{cases}
\end{equation}
Using this, one can extract $\rho_{D}(r)$ from the numerical eigenvalue density $\rho(x)$ from HMC simulations. We therefore have a prediction for any mixed moment at leading order at strong coupling. In figure \ref{fig:3dJointDensity}, we show the joint density for the $D=3$ SUSY model from the previous subsection. 

\section{Universality of huge operators}
\label{sec:universality}

We can deform a large $N$ matrix model by inserting different $U(N)$ invariants in the integral. The simplest case would be to insert some light single traces like $\tr X_1^2X_2X_1\ldots$ with number of matrices in the trace not scaling with $N$. Such an insertion can be treated like a probe in the large $N$ limit and does not affect the large $N$ saddle point. On the other hand, an insertion that scales like $N^2$ strongly backreacts and changes the background because it can compete with the measure and the action which also scale in the same way. 

In a holographic quantum theory, such $O(N^2)$ deformations, dubbed Huge operators in \cite{Kazakov:2024ald, Anempodistov:2025maj} also backreact and change the bulk geometry. A typical huge operator is expected to be dual to a black hole in the bulk. \footnote{The argument goes as follows: the Eigenstate Thermalization Hypothesis states that a typical high energy state is well described by a thermal ensemble. Separately, we have the fact that the bulk dual of a thermal CFT is a black hole.} Furthermore, classical black holes obey the no-hair theorem and therefore probe correlators in the presence of typical huge operators should only depend on some coarse-grained features of the huge operators. We cannot distinguish different black hole microstates using only simple probes.

Exploring such universality is quite challenging in say 4D $\mathcal N=4$ Super Yang-Mills, particularly so because we need to work at strong coupling. In \cite{Guerrieri:2025ytx}, it was found that the Hoppe model discussed above is a simple toy model that also exhibits the universality of huge operators. In particular, for a large class of huge operators in the Hoppe model at strong coupling, \cite{Guerrieri:2025ytx} showed that the eigenvalue density of the matrices is uniquely determined up to shifts and rescalings (and also rotations for the uplift analogous to \eqref{eq:rho4Defn}):
\begin{equation}
    \rho_{\texttt{huge}}(x) = \frac{L^2 - (x-x_0)^2}{4L^3/3}
\end{equation}
where the parabolic shape is universal at strong coupling but the size of the support $L$ and the center of mass depend on the details of the huge operator. Therefore, once we fix a few numbers, all probe correlators are determined. 
\begin{figure}
    \centering
   \includegraphics[width=\textwidth]{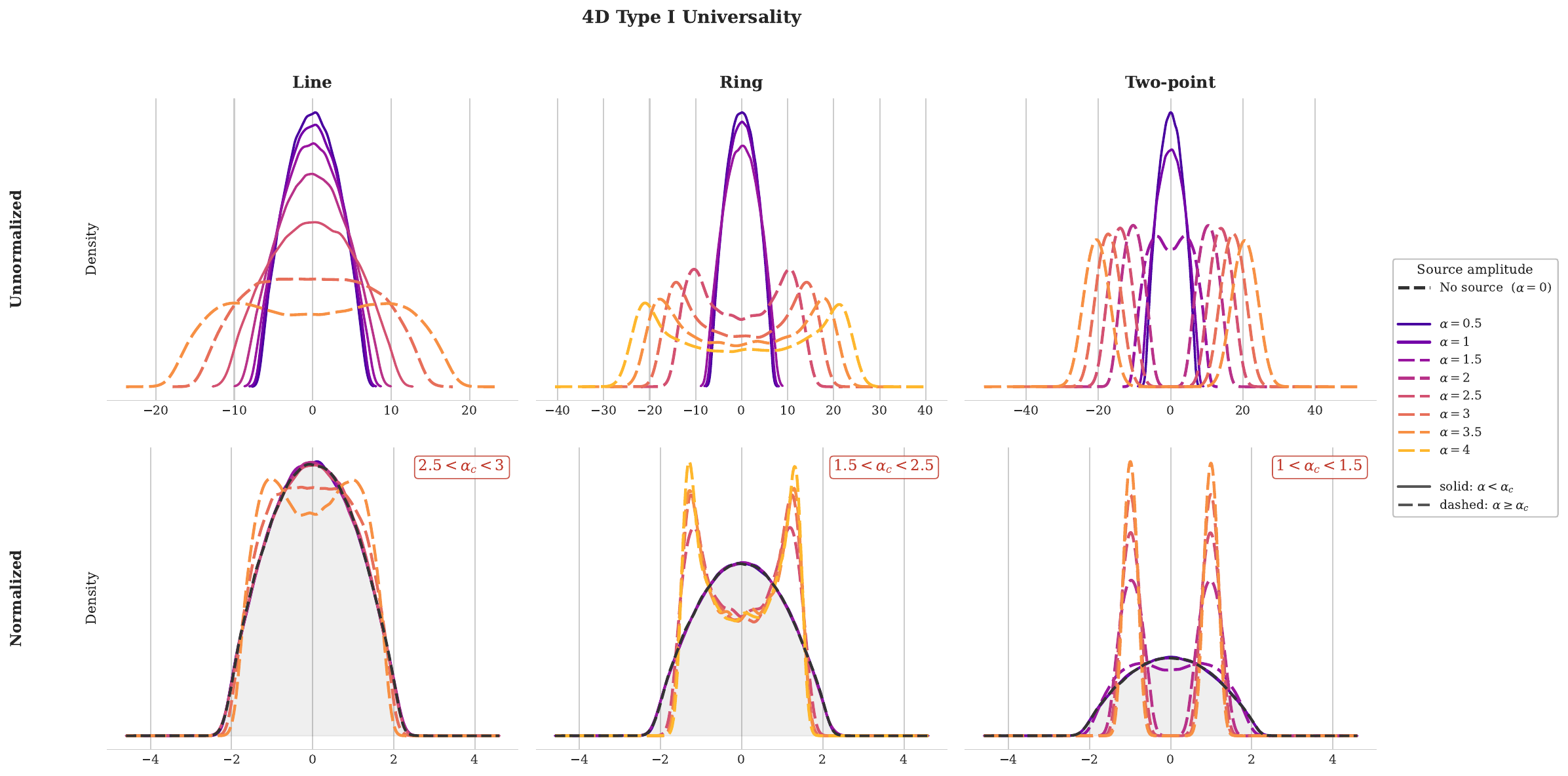}
    \caption{Universality of the eigenvalue distribution in the 4D type I model under different huge operators, at $N=50$ and $\lambda=150$. \textbf{Top:} Raw eigenvalue densities for the line, ring, and two-point source at various values of the source amplitude $\alpha$. \textbf{Bottom:} The same densities normalized by their standard deviation $\sigma$. For $\alpha < \alpha_c$ (solid lines), all three source geometries collapse onto the same universal distribution, which coincides with the normalized vacuum density (no source, dashed). Beyond $\alpha_c$, we are in a different phase where the distribution is non-universal and source-dependent. This mirrors the universal-to-non-universal transition found analytically in the Hoppe model~\cite{Guerrieri:2025ytx}.}
   \label{fig:type1Universality}
\end{figure}

Also, as one tunes the source there exists a phase transition from this universal phase to a non-universal phase. In Appendix \ref{app:hoppe} we review a particular simple example of a huge operator and the universal and non-universal phases in the Hoppe model. A natural question that arises is: since the $D=3$ and $D=4$ models studied above are commuting like the Hoppe model, do they also exhibit universality under different huge operators? While we cannot solve these models with sources analytically like for Hoppe, we can perform HMC simulations and the answer seems to be yes. 

Let us take the huge operator to be a classical source for the matrices
\begin{equation}
    O_{\texttt{huge}} = \exp\left(\frac{N}{\sqrt{\lambda}}\,\sum_I\tr\, J_I X_I\right)
\end{equation}
where the source matrices $J_I$ can be taken to have $O(1)$ eigenvalues such that the whole operator scales as $N^2$. The prefactor $\frac1{\sqrt\lambda}$ is chosen so as to have the universal to non-universal transition occur at a finite value of the source amplitude~\footnote{This is the same choice as in \cite{Guerrieri:2025ytx} after rescaling $X \to \sqrt{\lambda} X$.}. Let us consider three types of sources:
\begin{enumerate}
    \item Line source: We take $J_1$ to be a diagonal matrix $(J_1)_{ij} = \delta_{ij}\left(-\alpha + \frac{2j\alpha}N\right)$. That is, the diagonal entries of $J_1$ are uniformly distributed on a line of size $2\alpha$. The other source matrices are turned off, $J_{2\ldots4}=0$.
    \item Ring source: We take $J_1$ and $J_2$ to be diagonal with entries $(J_1)_{jj} = \alpha \cos\left(\frac{2\pi j}N\right)$ and $(J_2)_{jj} = \alpha \sin\left(\frac{2\pi j}N\right)$. That is, the source is uniformly distributed on a ring of radius $\alpha$ in the $X_1$-$X_2$ plane. Here, we set $J_3=J_4=0$.
    \item Two-point source: Only $J_1$ is turned on, and it has half its eigenvalues at $-\alpha$ and the other half at $+\alpha$.
\end{enumerate}
We ran HMC numerics for the 4D Type I model at $\lambda = 150$ and $N=50$ for various values of the source parameter $\alpha$. The results are shown in figure \ref{fig:type1Universality}. In the top panel, we see that the eigenvalue densities change as we change $\alpha$. However, in the bottom panel, we show the normalized densities i.e. the density of $\frac{x}{\sigma}$ where $\sigma$ is the standard deviation of the data. Upon doing this, the densities of the small $\alpha$ sources (the solid lines) collapse to a universal distribution, which matches the normalized density of the ``vacuum'' i.e. without any sources. In all the cases, we also see that there is a critical $\alpha_c$ beyond which the distribution is non-universal, like in the Hoppe model \cite{Guerrieri:2025ytx}. We find the same behaviour of universality in the 3D model, so both the SUSY models studied in section \ref{sec:susyModels} exhibit strong-coupling commutativity and huge-operator universality.

\section{Fermion number and the commutativity transition}
\label{sec:numFermions}
\begin{figure}[t]
    \centering
    \includegraphics[width=\linewidth]{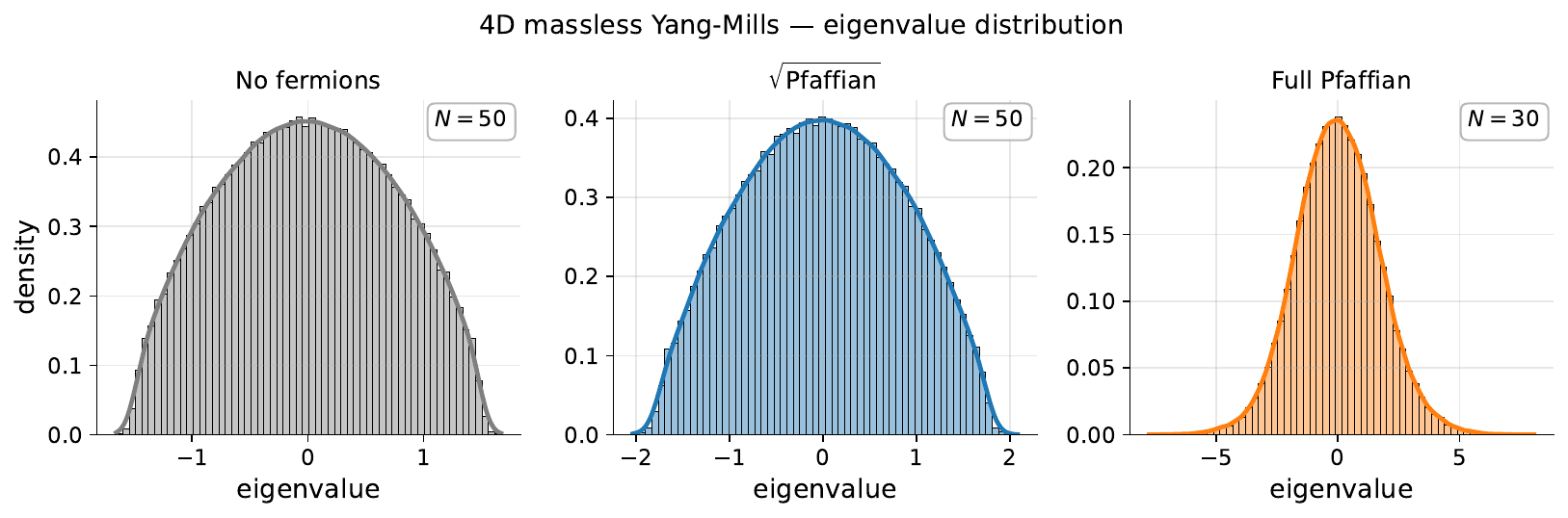}
    \caption{Eigenvalue density for the massless 4D model with different number of fermion flavours $\mathsf f=0, \frac12, 1$. For the bosonic model $\mathsf f=0$ and for $\mathsf f=\frac12$ the density has finite support and all moments are finite. The SUSY model $\mathsf f=1$ has power law tails and no finite positive moments for 4D. }
    \label{fig:massless4DPlots}
\end{figure}
The SUSY breaking examples of section \ref{sec:susyModels} suggest that supersymmetry is not the essential ingredient for commutativity. What matters more directly is the number of fermionic degrees of freedom. To vary this number in a controlled way, we take the fermionic Pfaffian in a given model and raise it to a power,
\begin{equation}
    \text{Pf}[\mathcal M(X)]\ \longrightarrow\ \left(\text{Pf}[\mathcal M(X)]\right)^{\mathsf f},
\end{equation}
where $\mathsf f$ can be analytically continued away from integer values. For the 4D Type~I model, $\mathsf f=1$ is the supersymmetric theory studied in section \ref{sec:susyModels}. This gives a simple way to probe whether the transition to commuting matrices is controlled by supersymmetry or by fermion number itself.

We find the following phase diagram for the strong coupling limit, with three different regimes. With too few fermions, the corresponding massless integral is finite, all moments are well defined, and the strong coupling limit of the mass deformed theory simply approaches this massless model. Since the massless model has no tunable coupling after rescaling the matrices, there is no reason for it to become commuting, and it does not. At the critical fermion number $\mathcal N_c=2(D-2)$, the massless model is at the edge of convergence. In particular, for the massless supersymmetric models where we have exactly these many fermions, the eigenvalue densities are expected to have power-law tails \cite{Krauth:1999qw}
\begin{equation}
    \rho(x) \sim |x|^{-2D+5}\quad,\qquad \text{for }|x|\gg1 ,\label{eq:susyPowerLawTail}
\end{equation}
so only the moments $\langle\tr X^{2k}\rangle$ with $k<D-3$ are finite. The (phase quenched) $D=3$ massless model is conjectured to be divergent, while in $D=4$ all positive moments diverge \cite{Krauth:1999qw}. Mass deforming these critical models leads instead to a commuting strong-coupling phase with a non-trivial, decaying scaling exponent for the commutativity ratio. Finally, above the critical fermion number the models are again commuting at strong coupling, but the limiting theory is different: as discussed in the next subsection, it is described by an effective theory of diagonal modes and the commutativity ratio obeys $R\sim\lambda^{-1}$ for all $D$.

\begin{figure}
    \centering
    \includegraphics[width=0.5\linewidth]{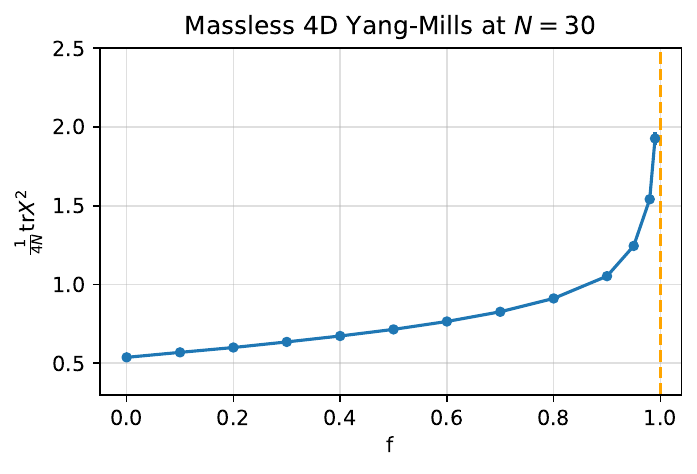}
    \caption{HMC results for the normalized second moment 
\(\frac{1}{4N}\langle \tr X_I X_I\rangle\) in the massless 4D Type~I model at \(N=30\), as a function of the Pfaffian power \(\mathsf f\). The moment remains finite for the subcritical models \(\mathsf f<1\), including the bosonic model \(\mathsf f=0\) and the intermediate model \(\mathsf f=\frac12\), but grows rapidly as \(\mathsf f\to1\). This signals the onset of the critical supersymmetric point, where the eigenvalue density develops a power-law tail and positive moments diverge in four dimensions.}
    \label{fig:momVsFerms}
\end{figure}

Let us first look at the subcritical side of this transition in the 4D Type~I model. Figure~\ref{fig:massless4DPlots} shows the eigenvalue density of the massless 4D model for $\mathsf f=0,\frac12$ and $1$. The bosonic model $\mathsf f=0$ has finite support, while the supersymmetric point $\mathsf f=1$ has a power-law tail. The intermediate model with $\mathsf f=\frac12$ behaves like the bosonic model: the eigenvalue density has finite support, and the moments are finite. In figure \ref{fig:momVsFerms}, we see that $\langle \tr X_IX_I\rangle$ is finite for the subcritical massless models, and diverges at the critical point $\mathsf f=1$ as expected from the power law tail \eqref{eq:susyPowerLawTail}.

\begin{figure}
    \centering
    \begin{subfigure}[c]{0.48\textwidth}
        \centering
        \includegraphics[width=\textwidth]{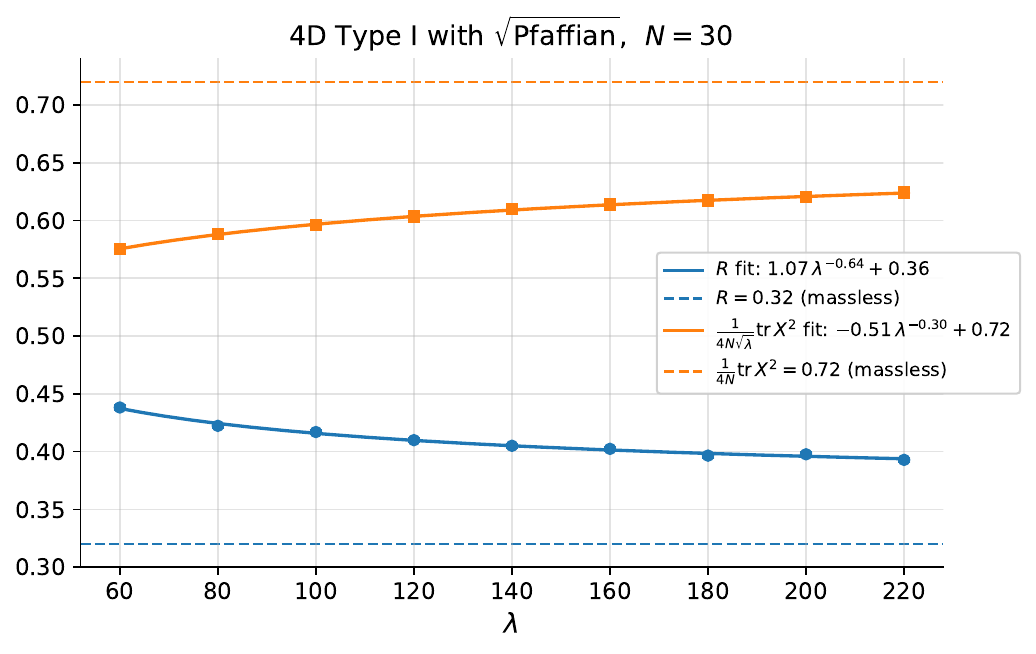}
    \end{subfigure}
    \hfill
    \begin{subfigure}[c]{0.48\textwidth}
        \centering
        \includegraphics[width=\textwidth]{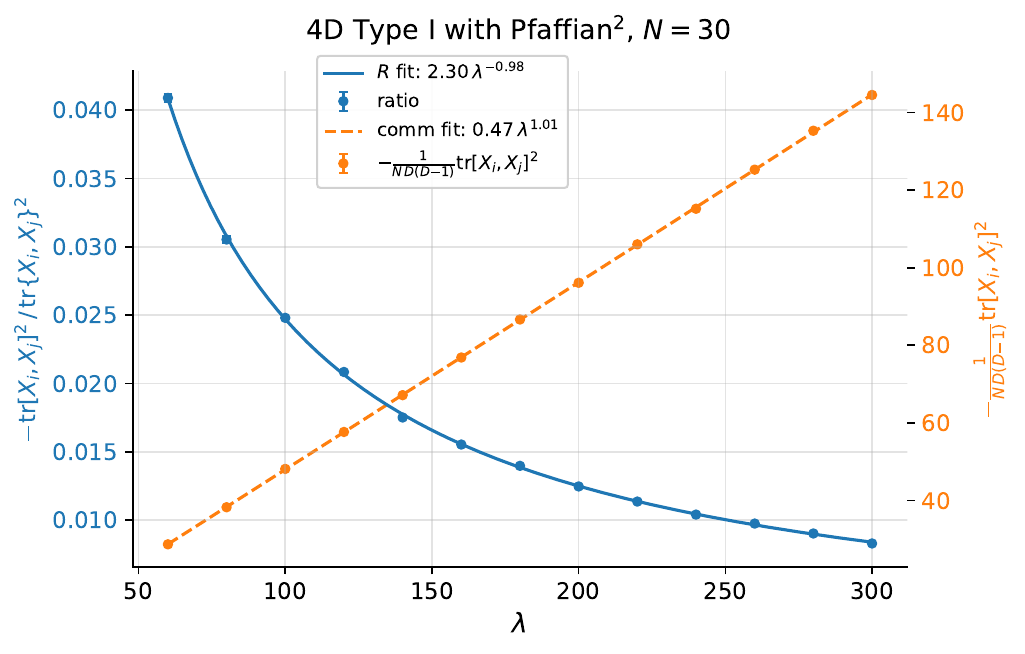}
    \end{subfigure}
    \caption{Monte Carlo results for the subcritical and supercritical cases of $\mathsf f=\frac12$ and $\mathsf f=2$. For the subcritical case, the strong coupling is non-commutative and all moments asymptote to their massless-model values. For the supercritical $\mathsf f=2$, the commutativity ratio scales as $R\sim\frac1\lambda$ and therefore is commuting.}
    \label{fig:halfdet_typeI}
\end{figure}

Let us now mass deform the $\mathsf f=\frac12$ model. Repeating the HMC analysis, we find the same behavior as in the $D\geq3$ bosonic Yang-Mills models discussed in the introduction. As shown in figure \ref{fig:halfdet_typeI}, the commutativity ratio does not decay to zero, but instead approaches a finite value $R\sim0.36$, in agreement with the value measured directly in the massless model up to statistical errors. Similarly, all moments converge to their massless-model values. Thus, with $\mathsf f=\frac12$, there are not enough fermions to overcome the entropic suppression of the commuting valleys. This is the subcritical phase of the transition.

We can then turn up the fermion number even further by going to $\mathsf f=2$. Since the supersymmetric point $\mathsf f=1$ is already barely convergent without mass, the massless model with $\mathsf f=2$ diverges. With a mass deformation, and going to the strong coupling limit, we find that the model is commuting with the commutativity ratio scaling as $R\sim \frac1\lambda$. As we will see below, the $\frac1\lambda$ scaling and other correlators in the supercritical regime can be computed in an effective theory of diagonal modes.

\subsection{Massive adjoint QCD models}
To further probe the supercritical side of the transition without the awkward analytic continuation, we now discuss a matrix model which naturally has supercritical fermionic degrees of freedom. We can do this while preserving $SO(D)$ by relaxing the supersymmetric reality and chirality constraints i.e. using Dirac fermions. This gives a matrix integral that we will call the mass deformed adjoint QCD matrix model,
\begin{equation}
    S_{\texttt{adjoint-QCD}} = \frac N\lambda\tr\left(-\frac14 [X_I,X_J]^2 - i\bar\psi\Gamma^I[X_I,\psi] + X_I^2 + \bar\psi\psi\right)
\end{equation}
where $\bar\psi$ and $\psi$ are independent Grassmann matrices and $\Gamma^I$ are Euclidean gamma matrices, which can be taken to be Hermitian. This model is defined in any dimension $D$ and preserves $SO(D)$. The Dirac spinor has $2^{\lfloor D/2\rfloor+1}$ real degrees of freedom, which is larger than the critical value $2(D-2)$ for all dimensions. We therefore expect these models to lie on the supercritical side of the transition.

After integrating out the fermions, we obtain the determinant
\begin{equation}
    \det\left(\mathbb{1}_{n_{\psi}(N^2-1)} - \sum_I \Gamma^I \otimes\mathbf{X}_I\right)\label{eq:detAdjointQCD}
\end{equation}
where $n_\psi = 2^{\lfloor\frac D2\rfloor}$ and $(\mathbf X_I)_{AB}=f_{ABC}X_I^C$ acts in the adjoint. The matrix $\sum_I\Gamma^I\otimes\mathbf{X}_I$ is anti-Hermitian, so the eigenvalues of \eqref{eq:detAdjointQCD} are generically of the form $1+i\mu_j$. The determinant is therefore complex in general.

In even dimensions $D=2m$, however, the adjoint QCD model has no sign problem. The chiral matrix $\Gamma_*=(-i)^m\prod_{I=1}^{2m}\Gamma^I$ anti-commutes with all the gamma matrices and squares to the identity. Therefore
\begin{equation}
    \Gamma_* \left(1-\sum_I \Gamma^I\otimes \mathbf{X}_I\right) \Gamma_*^{-1} = 1+\sum_I \Gamma^I\otimes \mathbf{X}_I
\end{equation}
which pairs every eigenvalue $1+i\mu_j$ with $1-i\mu_j$ and makes the determinant positive.

Note that the fermionic determinant in the $D=3$ and $D=4$ adjoint QCD models is the square of the Pfaffian of the corresponding supersymmetric models\footnote{In $D=3$, we mean $|\text{Pfaffian}|^2 = |\text{Det}|$}. Thus, in four dimensions, adjoint QCD is just the $\mathsf f=2$ version of the 4D Type~I model that we already encountered in the previous subsection. 

Interestingly the $D=5$ model is also free from the sign problem\footnote{In 5D, we can define the matrix $\mathcal C=\Gamma_2\Gamma_5$ with $\Gamma_5 = -\Gamma_1\ldots\Gamma_4$ (using the 4D gamma matrix conventions as in \cite{Martina:2025kwc}). With this, we have $\mathcal C \Gamma_I \mathcal C^{-1} = -\Gamma_I^*$ which ensures that the eigenvalues pair up to give a positive determinant.}. Another important observation is that we do not need to have supersymmetric field content in order to reach critical fermion number. For instance, in $D=5$ we have $\mathcal N_c=6$ and there is no way to start with the $2^{\lfloor D/2\rfloor+1}=8$ dimensional Dirac spinor and reduce to 6 real components. However, we can simply raise the fermionic determinant of the 5D adjoint QCD model to the power $\mathsf f=3/4$ to achieve this.

\begin{figure}
    \centering
    \begin{subfigure}[c]{0.48\textwidth}
        \centering
        \includegraphics[width=\textwidth]{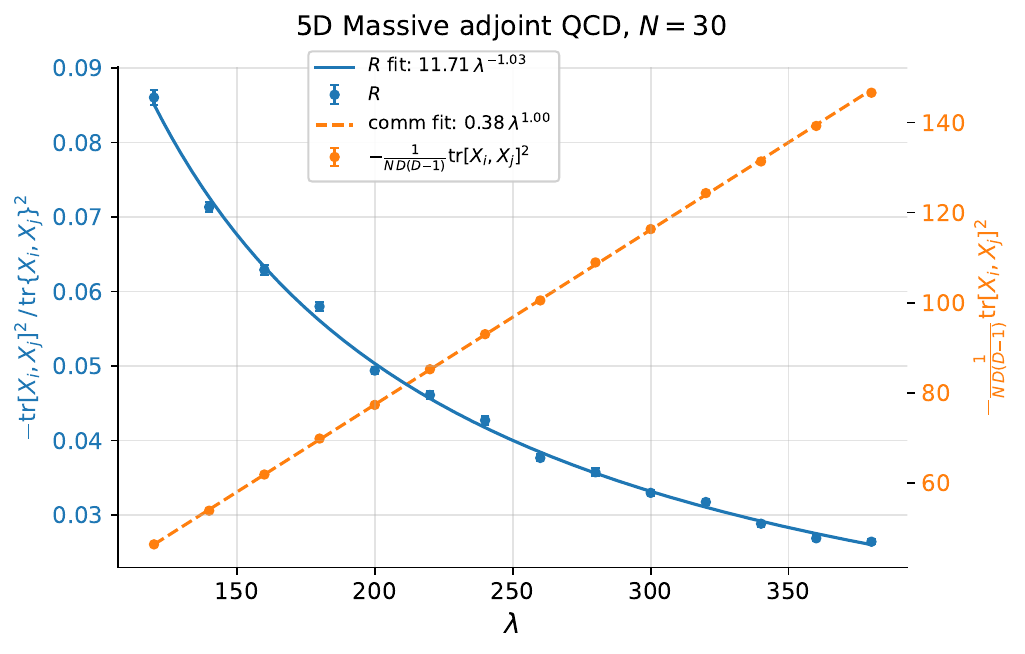}
    \end{subfigure}
    \hfill
    \begin{subfigure}[c]{0.48\textwidth}
        \centering
        \includegraphics[width=\textwidth]{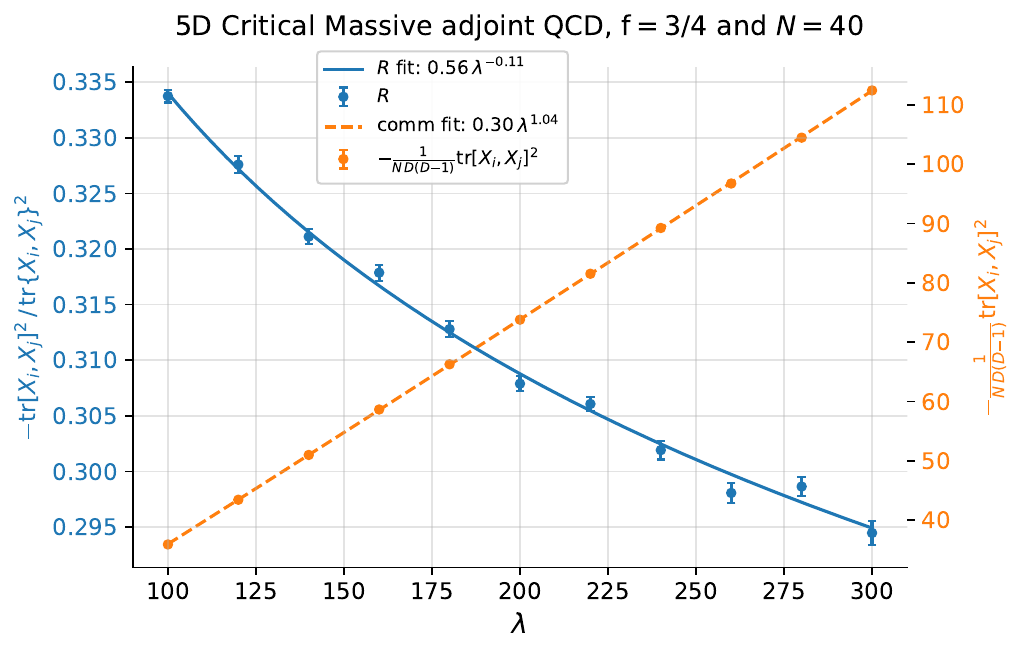}
    \end{subfigure}
    \caption{Commutativity ratio $R$ vs.\ $\lambda$ for the supercritical (left) and critical (right) 5D adjoint QCD models. For the supercritical case, we have the full fermionic determinant of size $4(N^2-1)\times4(N^2-1)$. We find that the commutativity ratio obeys $R\sim\frac1\lambda$ just like the 4D supercritical example. The critical case is obtained by raising the fermionic determinant of 5D adjoint QCD to a $\frac34$ power. For this case, we have a critical exponent of $0.11$.}
    \label{fig:adjQCD}
\end{figure}

In figure \ref{fig:adjQCD}, we show the HMC results for $D=5$ for both the adjoint QCD model and its critical counterpart with $\mathsf f=\frac34$. For the supercritical case, we find that the commutativity ratio scales as $R\sim \frac1\lambda$, whereas the critical model has the very different non-trivial scaling exponent of $0.11$.

The main lesson from these explorations is that mass deformations of singular or nearly singular massless models behave very differently from mass deformations of finite massless models. In the subcritical case, the strong coupling limit simply approaches the finite massless integral and remains non-commuting. At and above the critical fermion number, the massless limit is singular in the sense relevant for the partition function or for sufficiently high moments. The mass term regulates this singularity, and the resulting strong-coupling theory becomes commuting. The critical models show the nontrivial exponents discussed above, while the supercritical models have the simpler scaling $R\sim\lambda^{-1}$ and are described by a diagonal effective theory in the next section.

\section{Effective theory for the diagonal modes}
\label{sec: effective theory}
At sufficiently large coupling, and for configurations in which the eigenvalues are well separated, the off-diagonal modes acquire large masses and can be integrated out perturbatively. This leads to an effective theory for the diagonal modes. In this section we derive this effective description, discuss when it is self-consistent, and distinguish the subcritical, critical, and supercritical regimes:
\[
    \mathcal N<\mathcal N_c,\qquad
    \mathcal N=\mathcal N_c,\qquad
    \mathcal N>\mathcal N_c,
\]
where \(\mathcal N\) denotes the number of real fermionic components per adjoint generator and
$\mathcal N_c=2(D-2)$ as before. Consider a Yang-Mills matrix model with the action
\al{
\spl{
S = \frac{N}{\lambda}\mathrm{tr} \Biggl[ -\frac{1}{4} & [X_I,X_J]^2 - \frac{i}{2} \bar{\psi}\Gamma^{I} [X_I,\psi]  + X_{I} X_I + \bar \psi \psi \Biggr] \,. \label{eq: generic YM action}
}
}
This encompasses all the matrix models studied in this paper, upon interpreting $\psi$ as either a minimal spinor, a Dirac spinor, or $\mathsf f$ such spinors where $\mathsf f$ can be interpreted as the power to which the fermionic Pfaffian is raised. We will denote by $\mathcal{N}$ the total number of real fermionic d.o.f. 

We split the matrices into diagonal and off-diagonal parts,
\begin{equation}
    X_I = r_I + q_I \,, \qquad \psi_\alpha = \theta_\alpha + \Theta_\alpha\label{eq:changeOfVarsOffdiag}
\end{equation}
where $r_I, \theta_\alpha$ are diagonal matrices and $q_I, \Theta_\alpha$ are off-diagonal matrices. Using $SU(N)$ transformations, one can fix $(N^2-N)$ degrees of freedom among these variables. For example, one could diagonalize $X_D$, setting $q_D = 0$. Here however, we use the $SO(D)$ preserving constraint $\sum_I[r_I, q_I] = 0$. To make the strong-coupling expansion manifest, we also define rescaled coordinates 
\begin{equation}
    r_I \equiv \sqrt{\lambda} x_I, \qquad \theta_\alpha \equiv \sqrt{\lambda} \eta_\alpha \,, \qquad \Theta_\alpha = \lambda^{1/4} \chi_\alpha \,.
\end{equation}
In these coordinates, the action reduces to
\begin{align}
\begin{split}
    S & = N \Bigl[x_a^I x_a^I + \bar \eta_a^\alpha \eta_a^\alpha + \frac{1}{2}  |x_{ab}|^2 |q_{ab}|^2 - \frac{i}{2} x_{a b}^I \bar \chi_{ba} \Gamma^I \chi_{ab} \Bigr]   + S_\mathrm{higher}  \label{eq: S expanded}
\end{split}
\end{align}

where $x_{a}^I$ are the components of the diagonal matrix $x^I$, $x_{ab}^I \equiv x_a^I - x_b^I$, and $|\cdot|$ denotes the $D$-dimensional Euclidean norm. 

The terms collected in $S_\mathrm{higher}$ are suppressed, at fixed $x_{ab}$, by powers of $\lambda^{-1/4}$. They include, schematically, interactions such as $xq^3/\sqrt{\lambda}$ and $q^4/\lambda$.
A more explicit expansion, together with the full Faddeev-Popov gauge fixing determinant, is given in \eqref{eq: part func in SO(D) preserving gauge}. The diagonal approximation is therefore an expansion at fixed separations in the rescaled variables. It may break down if the dominant configurations have parametrically small $|x_{ab}|$, and this possibility will be crucial in comparing different values of $\mathcal{N}$ below.

Assuming $S_\mathrm{higher}$ can be dropped, we integrate out the off-diagonal bosonic and fermionic components using only the Gaussian terms, yielding
\begin{align}
\begin{split}
    \int [d q_I] [d \chi_\alpha] & \delta(\hat x_{ab}^I q_{ab}^I) \prod_{a < b} |x_{ab}|^2 \exp\left(- \frac{N}{2} |x_{ab}|^2 |q_{ab}|^2 + \frac{i N}{2} x_{ab}^I \bar \chi_{ba} \Gamma^I \chi_{ab}\right) \\ & \qquad \qquad \propto \prod_{a < b} |x_{ab}|^2 |x_{ab}|^\mathcal{N} |x_{ab}|^{- 2(D-1)} = \prod_{a < b} |x_{ab}|^{\mathcal{N} - \mathcal{N}_c} \,.
\end{split}
\end{align}
The factor $|x_{ab}|^\mathcal{N}$ comes from the $\mathcal{N}$ off-diagonal fermions, the factor $|x_{ab}|^{-2(D-1)}$ comes from the $D-1$ off-diagonal bosons \footnote{Note that one is killed by the constraint $\hat x_{ab}^I q_{ab}^I = 0$.}, and $|x_{ab}|^{2}$ comes from the leading term in the Faddeev-Popov determinant.

Returning to the original coordinates $r_I$, the effective theory for the bosonic diagonal modes thus reads
\begin{equation}
    \langle f(X_I) \rangle \xrightarrow{\lambda \to \infty} \langle f(r_I) \rangle_\mathrm{diag}  \,,\label{eq: observables massless limit}
\end{equation}
where
\begin{equation}
    \langle f(r_I) \rangle_\mathrm{diag} \equiv \frac{\int\prod_{a=1}^N \left( \prod_{I=1}^D dr^I_a \right) f(r_I) \ \delta^{(D)}\left( \sum_a \vec r_a\right)\prod_{a < b} |r_{ab}|^{\mathcal{N} - \mathcal{N}_c} \ \mathrm{exp}\left(- \frac{N}{\lambda} \sum_a|r_a|^2\right)}{\int\prod_{a=1}^N \left( \prod_{I=1}^D dr^I_a  \right) \ \delta^{(D)}\left( \sum_a  \vec r_a\right)\prod_{a < b} |r_{ab}|^{\mathcal{N} - \mathcal{N}_c} \ \mathrm{exp}\left(-\frac{N}{\lambda} \sum_a|r_a|^2\right)}\label{eq:diagonalMoments}
\end{equation}
and we recall that $\mathcal{N}_c \equiv 2(D-2)$.

The power of the Vandermonde factor leads to three qualitatively different cases. In the subcritical case $\mathcal{N}< \mathcal{N}_c$, the effective measure is singular when diagonal eigenvalues collide. If enough eigenvalues collapse, the integral in \eqref{eq:diagonalMoments} diverges; in particular, the full $N$-eigenvalue collision is divergent for
\begin{equation}
    N \geq \frac{2D}{\mathcal{N}_c-\mathcal{N}}\,.
\end{equation}
In this regime the diagonal theory is dominated by precisely the small-$|x_{ab}|$ region where the expansion that led to \eqref{eq:diagonalMoments} is not controlled. Thus, diagonal and off-diagonal modes are not separated in a parametrically reliable way, and one should not necessarily expect commutativity. This is consistent with the numerical results: the $D=3$, $\mathcal N=0$ model in Figure~\ref{fig:HoppeVs3D} and the $D=4$, $\mathcal N=2$ model in Figure~\ref{fig:halfdet_typeI} both approach a non-zero commutativity ratio.

In the critical case $\mathcal{N}=\mathcal{N}_c$, the induced Vandermonde factor is absent. Consequently, at fixed $N$ and in the limit $\lambda\to\infty$, the $SU(N)$ model reduces to a free $U(1)^{N-1}$ theory of traceless diagonal matrices. In terms of the rescaled variables, the typical pairwise separation behaves as $|x_{ab}|\sim N^{-1/2}$. For fixed $N$, these separations remain finite, and the strong-coupling expansion around the diagonal sector is therefore self-consistent. The situation is more subtle when $N$ and $\lambda$ are taken to infinity simultaneously: the diagonal separations shrink with $N$, and the validity of the approximation depends on how rapidly $\lambda$ grows relative to $N$. We will see numerically that, once $\lambda$ is taken far beyond the 't~Hooft window, the diagonal approximation indeed becomes reliable.

Finally, in the supercritical case $\mathcal{N}>\mathcal{N}_c$, the diagonal measure contains a genuine repulsion. 
In the rescaled variables $r_I=\sqrt{\lambda}x_I$, the diagonal effective action is
\begin{equation}
    S_\mathrm{eff}
    =
    N\sum_a |x_a|^2
    -(\mathcal N - \mathcal N_c) \sum_{a < b} \log |x_{ab}| \,.
    \label{eq: effective supercritical theory}
\end{equation}
The logarithmic repulsion keeps the eigenvalue separations at $|x_a|,|x_{ab}|=\mathcal O(1)$ even at large $N$. The diagonal approximation is therefore self-consistent already for $\lambda\gg1$, including the 't~Hooft regime where $N$ is parametrically larger than $\lambda$.

\subsection{Predictions and comparison to numerics}

The discussion above gives two uses of the diagonal theory. In critical models it describes the finite-$N$ ultra-strong coupling limit. In supercritical models it remains valid for the 't Hooft regime, $N\gg\lambda\gg1$. We now extract a few observables in these two cases and compare them with the numerical data.

\paragraph{In the critical case:} When $\mathcal{N} = \mathcal N_c$, we can directly use \eqref{eq:diagonalMoments} and obtain
\begin{equation}
    \frac{1}{N}\langle \mathrm{tr} X_1^2 \rangle \xrightarrow{\lambda \to \infty} \lambda \left(\frac{N-1}{2N^2} \right)\label{eq:trX2DiagonalApprox}
\end{equation}
for each matrix direction. The simulations shown in Figure~\ref{fig:type1-results} were performed at $N=50$ and $50<\lambda<450$, which is a 't~Hooft-regime scan rather than the regime where the critical diagonal theory is expected to be accurate. To test the fixed-$N$ strong-coupling prediction directly, Table~\ref{tab: Strong coupling of 4d type I} compares \eqref{eq:trX2DiagonalApprox} with simulations of the 4D Type~I model at much larger couplings. The agreement improves as $\lambda$ is increased and is within statistical errors around $\lambda\simeq19000$.
\begin{figure}
    \centering
    \includegraphics[width=\linewidth]{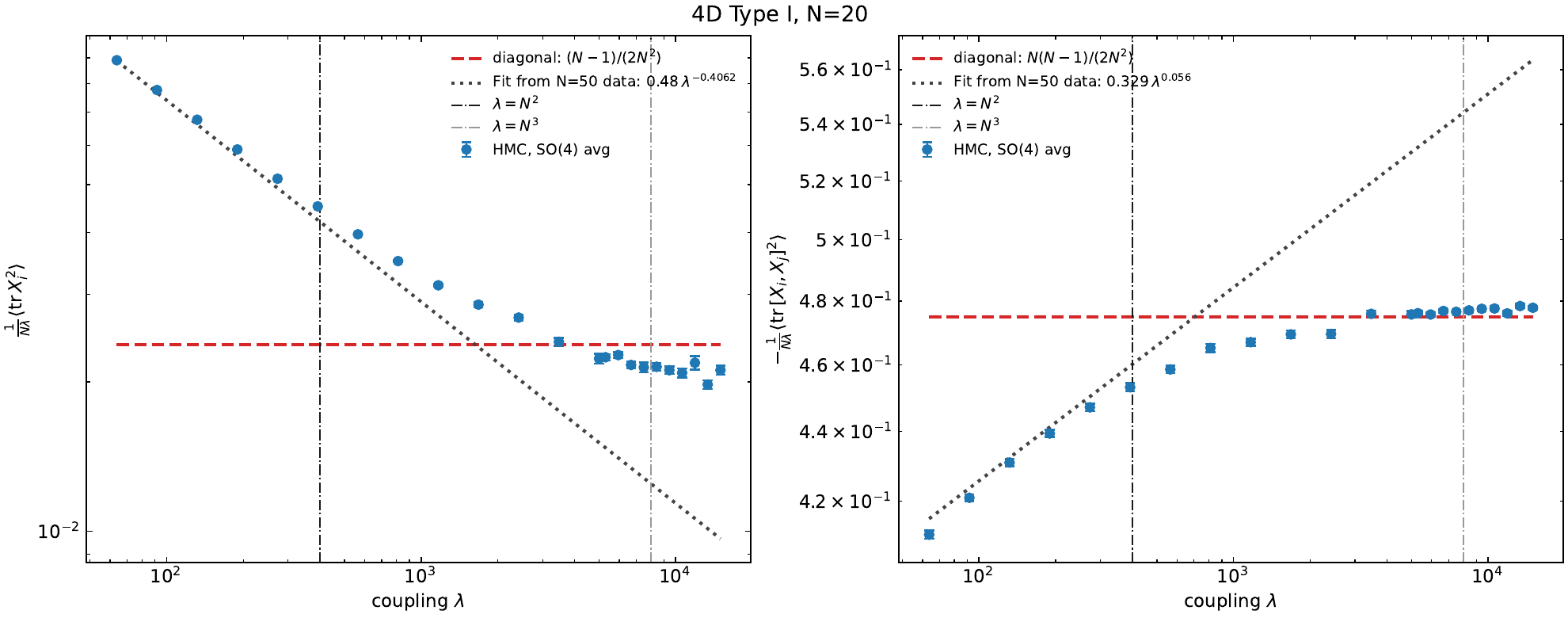}
\caption{
Crossover to the diagonal regime in the critical 4D Type~I model at \(N=20\).
At large \(\lambda\), the normalized second moment (left) and commutator (right)
depart from their 't~Hooft-regime fits (dotted) and approach the fixed-\(N\)
diagonal predictions (dashed). Vertical lines mark \(\lambda=N^2\) and \(N^3\).
}
\label{fig:typeI-ultrastrong}
\end{figure}

Figure~\ref{fig:typeI-ultrastrong} shows this crossover explicitly
for the critical 4D Type~I model at \(N=20\). As \(\lambda\) is
increased far beyond the 't~Hooft window, both
\(\frac{1}{N\lambda}\langle\tr X_i^2\rangle\) and
\(-\frac{1}{N\lambda}\langle\tr[X_i,X_j]^2\rangle\)
depart from the extrapolated 't~Hooft-regime power laws and approach
the constant values predicted by the fixed-\(N\) diagonal theory in
\eqref{eq:trX2DiagonalApprox} and \eqref{eq:comm2DiagonalApprox}.

\begin{table}[h]
\centering
\renewcommand{\arraystretch}{1.4}
\begin{tabular}{|c|c|c|c|c|c|}
\hline
$\lambda$
&
$11000$
&
$13000$
&
$15000$
&
$17000$
&
$19000$
\\
\hline
$\displaystyle 
\left\langle \operatorname{tr} X_1^2 \right\rangle_{\rm num}$
&
$6.3 \cdot 10^3$
&
$7\cdot 10^3$
&
$7.7 \cdot 10^3$
&
$8.4 \cdot 10^3$
&
$9.3 \cdot 10^3$
\\
\hline
$\displaystyle 
\left\langle \operatorname{tr} X_1^2 \right\rangle_{\lambda \to \infty}$
&
$5.390\cdot 10^3$
&
$6.370\cdot 10^3$
&
$7.350\cdot 10^3$
&
$8.330\cdot 10^3$
&
$9.310\cdot 10^3$
\\
\hline
\end{tabular}
\caption{Numerical values of $\langle \operatorname{tr} X_i^2\rangle$ for different values of $\lambda$ in 4D type I at $N=50$, and their comparison to the $\lambda \to \infty$ approximation}
\label{tab: Strong coupling of 4d type I}
\end{table}

\paragraph{In the supercritical case:} For $\mathcal N>\mathcal N_c$, the diagonal theory has a non-trivial and controlled large-$N$ limit. The empirical distribution of the diagonal vectors becomes a continuous density $\rho_{D}(\vec r)$ in the original $r$ variables. In Appendix~\ref{app:strong coupling fixed N} we derive that the large-$N$ saddle is supported on a $D$-dimensional ball for $D<4$, and on a $(D-1)$-dimensional spherical shell for $D\geq4$:
\begin{equation}
    D< 4 : \quad \rho_{D,\nu}(\vec r)
    =
    \frac{\left(R_D^2-|\vec r|^2\right)^{1-\frac D2}}
    {\pi^{D/2}\Gamma\left(2-\frac D2\right)R_D^2}\,
    \Theta(R_D-|\vec r|)\,,
    \qquad
    R_D^2=\frac{2\nu\lambda}{D}\,.
\end{equation}
\begin{equation}
    D\geq 4 : \quad \rho_{D,\nu}(\vec r)
    =
    \frac{\delta(|\vec r|-R_D)}{\Omega_{D-1}R_D^{D-1}}\,,
    \qquad
    R_D^2=\frac{\nu\lambda}{2}\,,
    \qquad
    \Omega_{D-1}=\frac{2\pi^{D/2}}{\Gamma(D/2)}\,,
\end{equation}
where $\nu = (\mathcal N - \mathcal N_c)/2$.
These densities determine all leading commutative single-trace observables in the sense of \eqref{eq:rho4Defn}: at this order the ordering of the matrices inside the trace is irrelevant.

The leading commutator is obtained from the Gaussian off-diagonal fluctuations in \eqref{eq: S expanded}. This gives
\begin{equation}
    -\frac{1}{N} \langle \mathrm{tr} [X_1, X_2]^2 \rangle \xrightarrow{\lambda \to \infty} \frac{2\lambda}{D} \frac{N(N-1)}{N^2} \label{eq:comm2DiagonalApprox}
\end{equation}
at fixed $N$. Combining this result with the large-$N$ diagonal density and factorization, one obtains
\begin{equation}
    R = \left\langle\frac{
    -\sum_{I<J}\tr[X_I,X_J]^2
    }{
    \sum_{I<J}\tr\{X_I,X_J\}^2
    }
    \right\rangle \xrightarrow[N\to\infty]{\lambda\to\infty} \begin{cases}
        \frac{3D}{\nu^2 \lambda} , & D<4 \\ \frac{2(D+2)}{\nu^2 \lambda} & D \geq 4
    \end{cases}
\end{equation}

Let us compare these results to the numerical fits from previous sections for the supercritical models. In tables \ref{tab: commsquared comparison} and \ref{tab: ratio comparison}, we report the results from the numerical fits of HMC, analytic predictions of the diagonal theory at large $N$ and also finite $N$ results obtained by numerically sampling from \eqref{eq:diagonalMoments}. It is clear that the supercritical cases of 4D and 5D adjoint QCD are very well described by the effective theory of diagonals. 

\begin{table}[h]
\centering
\renewcommand{\arraystretch}{1.4}
\begin{tabular}{|c|c|c|}
\hline
$-\frac{1}{N}
\left\langle \operatorname{tr}[X_1,X_2]^2 \right\rangle$
 & 
\begin{tabular}{c}
$4d$ adj QCD \\
$N=30$
\end{tabular}
&
\begin{tabular}{c}
$5d$ adj QCD \\
$N=30$
\end{tabular}
\\
\hline
\begin{tabular}{c}
Numerical fits
\end{tabular}
&
$0.47\,\lambda$
&
$0.38\,\lambda$
\\
\hline
\begin{tabular}{c}
Effective theory $\lambda \to \infty,\ N=30$
\end{tabular}
&
$0.483\,\lambda$
&
$0.386\,\lambda$
\\
\hline
Effective theory $\lambda \to \infty, \ N\to \infty$
&
$0.5\,\lambda$
&
$0.4\,\lambda$
\\
\hline
\end{tabular}
\caption{Comparison of the commutator between the numerical results which probe the 't~Hooft limit (top row) and the diagonal approximation at finite and large $N$ (bottom two rows respectively). }
\label{tab: commsquared comparison}
\end{table}

\begin{table}[h]
\centering
\renewcommand{\arraystretch}{1.4}
\begin{tabular}{|c|c|c|}
\hline
$\displaystyle 
\left\langle \frac{
-\sum_{I<J}\operatorname{tr}[X_I,X_J]^2 
}{
\sum_{I<J}\operatorname{tr}\{X_I,X_J\}^2 
}\right\rangle$
 & 
\begin{tabular}{c}
$4d$ adj QCD \\
$N=30$
\end{tabular}
&
\begin{tabular}{c}
$5d$ adj QCD \\
$N=30$
\end{tabular}
\\
\hline
\begin{tabular}{c}
Numerical fits
\end{tabular}
&
$2.3\,\lambda^{-0.98}$
&
$11.71\,\lambda^{-1.03}$
\\
\hline
Effective theory $\lambda \to \infty,\ N=30$
&
$2.52\,\lambda^{-1}$
&
$9.6\,\lambda^{-1}$
\\
\hline
Effective theory $\lambda \to \infty,\ N\to \infty$
&
$3\,\lambda^{-1}$
&
$14\,\lambda^{-1}$
\\
\hline
\end{tabular}
\caption{Comparison of the ratio $R$ between HMC numerical fits and diagonal theory. The ratio amplifies the statistical errors and which is why the mismatch is larger here than in table \ref{tab: commsquared comparison}}
\label{tab: ratio comparison}
\end{table}

This diagonal effective theory also explains why we do not expect huge-operator universality in the supercritical models. Once the off-diagonal modes have been integrated out, the remaining coupling can be removed by rescaling the diagonal variables as shown in \eqref{eq: effective supercritical theory}. A huge $O(N^2)$ source then enters directly in the saddle point equations for the vectors $x_a$, and different microscopic choices of the source lead to different diagonal distributions. We show this explicitly with some deformation examples in appendix \ref{app: non-universal supercritical}. Thus, the supercritical models turn out to be commuting without having the no-hair-like universality seen in the critical examples.

\section{Discussion}

In this paper, we have explored the strong coupling limit of mass deformed Yang-Mills matrix models in various dimensions and with different numbers of fermions. The main driving question was to determine when the strong-coupling theory becomes effectively commuting, so that the eigenvalues can potentially be viewed as coordinates of an emergent commutative geometry. Our numerical results point to a simple organizing principle: the number of fermionic degrees of freedom controls the competition between the Yang-Mills attraction toward commuting valleys and the entropic suppression of those valleys. With too few fermions, the entropic effect wins and the model remains non-commutative at strong coupling; the bosonic Yang-Mills models with $D\geq3$ are in this class. At the critical value $\mathcal{N}_c=2(D-2)$, the massless model sits at the edge of convergence, and in the models we studied, its mass deformation instead flows to a commuting strong-coupling phase. In $D=3,4,6$ and $10$, this critical value is precisely the fermion content of supersymmetric Yang-Mills. Above this threshold the models are also commuting, but the nature of the strong-coupling limit changes, as discussed below. The scaling exponents for the commutativity ratio for critical models are shown in table \ref{tab:critExponents}; within the dimensions we studied, they appear to decrease monotonically with $D$.

The diagonal effective theory obtained by integrating the off-diagonals provides a simple description of the leading strong-coupling behavior, and its predictions agree well with our numerical results for critical models once the coupling is sufficiently large. However, doing a systematic expansion around the diagonal theory is considerably more subtle. Configurations in which diagonal eigenvalues approach one another lead to naive divergences that need to be regulated and invalidate the expansion in off-diagonal fluctuations. A systematic treatment requires a resummation of diagrams associated with these near-collision regions. Establishing this resummation, and thereby determining the phase structure and crossover between the ’t Hooft and diagonal regimes in the critical models, remains an open problem.

\begin{table}
    \centering
    \begin{tabular}{|c|c|}
        \hline
        Dimension & Exponent in $R\sim\lambda^{-\alpha}$ \\
        \hline
        2 & $\frac13$\\
        3 & 0.25\\
        4 & 0.16\\
        5 & 0.11\\
        \hline
    \end{tabular}
    \caption{Critical exponents of the commutativity ratio in various dimensions in the case of $SO(D)$ preserving models. The $1/3$ in $D=2$ is analytically known.}
    \label{tab:critExponents}
\end{table}

We conjecture that the mass deformed models with critical fermion number exhibit a no-hair-like universality: in the presence of a typical $O(N^2)$ insertion, light probes are sensitive only to coarse-grained data of the huge operator, rather than to its microscopic details. We checked this explicitly for the 4D Type~I and 3D SUSY models, and the Hoppe model provides another example in the same critical class. The supercritical models show that commutativity alone is not sufficient for this universality. Their strong coupling limit is commuting, and is described by an effective theory of diagonal modes even in the {}'t~Hooft limit, but this diagonal theory remains sensitive to the detailed form of the huge deformation.

It would be interesting to better understand concretely the relation between a commutative bulk eigenvalue space and locality in the emergent space. A good starting point is to study connected correlators in the solvable Hoppe model at strong coupling \cite{MuraliVieira:toappear}. The usual dictionary in holography relates single trace operators to strings in the bulk. So the connected correlators would encode information about geodesic distances in the putative bulk. This would provide a sharper test of whether the commuting eigenvalue space could be interpreted as a genuine bulk space, or if it is a suggestive but imprecise notion of a bulk geometry.

The matrix integrals studied in this paper do not have a manifest string-theoretic embedding. Their closest string-theoretic analogue is the polarized IKKT matrix model. In that model, however, fuzzy-sphere saddles introduce an additional complication, since their stability in the ’t Hooft limit is not guaranteed. We will return to polarized IKKT, and to the role of fuzzy spheres, in a companion paper.

We have only studied zero-dimensional matrix integrals in this work, so an important open question is whether the same mechanism survives in more established holographic theories. A natural next target is D0-brane matrix quantum mechanics \cite{Banks:1996vh,Itzhaki:1998dd}. The BFSS model has several features reminiscent of the massless supersymmetric matrix integrals studied here. Its ground-state wavefunction has a power-law tail, and only sufficiently low moments of the matrices are finite; in particular $\langle\tr X^k\rangle$ is finite only for $k<9$ \cite{Polchinski:1999br}. At finite temperature, the partition function is also argued to be divergent because of the gapless spectrum and the non-compact moduli space associated with arbitrarily separated D0 branes \cite{Kabat:1999hp,Kabat:2000cv,Catterall:2007fp}. These infrared issues can be regulated by adding a mass deformation. The most important example is the BMN model \cite{Berenstein:2002jq}, which preserves supersymmetry while lifting the flat directions.

In \cite{Komatsu:2024vnb}, it was shown that, at finite $N$ and in the strong-coupling limit, the BMN model expanded around the trivial vacuum reduces to a collection of free harmonic oscillators associated with the diagonal modes, and in particular $R \xrightarrow[]{\lambda \to \infty} 0$. This closely parallels our results for critical matrix models. However, as illustrated for instance by the Hoppe model, this limit can differ substantially from the 't Hooft limit. It would therefore be very interesting to study the strong 't~Hooft limit and how it differs from the finite $N$ strong coupling limit.

There are several ways to address this question. The ground state could be studied using variational machine-learning wavefunctions \cite{Han:2019wue,Bodendorfer:2024jti}, while the thermal ensemble can be accessed by Monte Carlo methods \cite{Asano:2018nol,Bergner:2021goh,Pateloudis:2022fyl,Jha:2024bue}. Another promising possibility is the matrix bootstrap approach \cite{Han:2020bkb,Lin:2023owt,Cho:2024kxn}. A bold extrapolation from the present results would suggest that the BMN model at strong coupling, or equivalently low mass, should belong to the commuting phase and exhibit universality under deformations by typical huge operators.
\paragraph{Acknowledgements}
We thank M. Guica, S. Hartnoll, M. Ligorio, J. Penedones, P. Vieira, A. Vuignier, X. Zhao for helpful discussions.  We also acknowledge the aid of artificial intelligence tools ChatGPT and Claude in the development of the HMC code. This work was supported by the Swiss National Science Foundation through the project 200020\_197160 and through the National Centre of Competence in Research SwissMAP. HM thanks ICTP-SAIFR for hospitality during part of this work. Research at Perimeter Institute is supported in part by the Government of Canada through the Department of Innovation, Science, and Economic Development Canada and by the Province of Ontario through the Ministry of Colleges and Universities.

\appendix

\section{Hybrid Monte Carlo}
\label{app:hmc}

All Monte Carlo data in this paper were generated using \texttt{matrix-hmc}, a publicly available\footnote{\url{https://github.com/harish02murali/matrix-hmc}} PyTorch-based HMC package for matrix models. Conceptually, the code mirrors the standard Monte Carlo workflow: define the Euclidean action, evolve with fictitious Hamiltonian dynamics, and accept/reject with a Metropolis step to maintain exact sampling. In practice, the implementation is compact and efficient because PyTorch provides fast linear algebra and automatic differentiation for force evaluation.

\subsection*{Quick usage}

The package can be installed from PyPI and used immediately from the command-line with existing built-in models:
\begin{lstlisting}[style=pycode, language=bash]
pip install matrix-hmc

matrix-hmc --model yangmills --nmat 4 --ncol 50 --coupling 150 \
    --niters 1000 --step-size 0.5 --nsteps 50 --name run --data-path data
\end{lstlisting}
For Python usage, the equivalent call is:
\begin{lstlisting}[style=pycode]
import matrix_hmc as hmc
from matrix_hmc.models.yangmills import YangMillsModel

hmc.configure(device="auto", precision="complex64")
model = YangMillsModel(dim=4, ncol=50, couplings=[150.0])
hmc.run(model, niters=1000, step_size=0.5, nsteps=50, output="data", name="run")
\end{lstlisting}

\subsection*{The algorithm}

The basic idea of Hybrid Monte Carlo (HMC) \cite{Duane:1987de} (see also the nice reviews in \cite{Hanada:2018fnp,Jha:2021exo}) is to introduce fictitious momenta $P_I$ conjugate to the matrix fields $X_I$ and to sample the extended distribution
\begin{equation}
    p(X, P) \propto e^{-H(X,P)}, \qquad H(X,P) = V(X) + \frac12 \sum_I \tr P_I^2,
\end{equation}
where $V(X)$ is the potential (the Euclidean action), and the $P_I$ are independent Gaussian random Hermitian matrices. One then proposes moves in $X$ by integrating Hamilton's equations
\begin{equation}
    \dot X_I = P_I, \qquad \dot P_I = -\frac{\partial V}{\partial X_I} \equiv -F_I(X),
\end{equation}
using a symplectic leapfrog integrator for $n_\text{steps}$ steps of size $dt$. The proposal is accepted or rejected via a Metropolis step with probability $\min(1, e^{-\Delta H})$, where $\Delta H = H_\text{final} - H_\text{initial}$. Since the leapfrog is symplectic and time-reversible, the algorithm is exact regardless of step size; $\Delta H \ne 0$ only due to numerical integration errors, which vanish as $dt \to 0$.

For fermionic matrix models, the effective potential after integrating out fermions is generically of the form
\begin{equation}
    V(X) = V_\text{bos}(X) - \tfrac12 \log \left|\det \mathcal{D}(X)\right|,
\end{equation}
where $\mathcal{D}(X)$ is a fermion operator of size $\mathcal{N}(N^2-1)\times \mathcal{N}(N^2-1)$ (with $\mathcal{N}$ the total number of real fermionic d.o.f.). The corresponding fermion force is the expensive part; formally,
\begin{equation}
    \frac{\partial}{\partial X_I}\log\left|\det\mathcal{D}\right| = \tr\Bigl[\mathcal{D}^{-1}\frac{\partial\mathcal{D}}{\partial X_I}\Bigr],
\end{equation}
which naively requires repeated large linear solves during each trajectory. For models where the determinant is retained, the code evaluates this contribution through \texttt{torch.linalg.slogdet} and obtains forces by autograd.

Running the HMC on the CPU at $N = 50$ with a Macbook M1 Pro is orders of magnitude slower than the Nvidia H200 GPU used for most of the plots in the paper (roughly $150$ times slower per force evaluation). 

\subsection*{Code structure}

The code is organized as follows:
\begin{itemize}
    \item \texttt{hmc.py} — model-agnostic HMC kernel: leapfrog trajectory generation and Metropolis accept/reject.
    \item \texttt{models/base.py} — abstract base class \texttt{MatrixModel}. Subclasses must implement \texttt{potential(X)} and \texttt{measure\_observables(X)}; the force $F_I = \partial V/\partial X_I$ is computed automatically via autograd and projected onto the Hermitian traceless subspace.
    \item \texttt{algebra.py} and \texttt{config.py} — linear algebra helpers and runtime/device configuration.
    \item \texttt{models/pikkt4d\_type1.py}, \texttt{pikkt4d\_type2.py}, \texttt{yangmills.py}, \ldots — existing model implementations.
\end{itemize}
The only interface the HMC engine requires from a model is \texttt{potential(X)}, \texttt{measure\_observables(X)}, and \texttt{get\_state()}/\texttt{set\_state(X)}. Everything else --- force computation, Hermitian projection, leapfrog jumps and acceptance tests --- is handled by the shared framework code.

\subsection*{Adding a new model}

Adding a new model requires only subclassing \texttt{MatrixModel} and implementing \texttt{potential}. As a minimal example, here is the massive bosonic Yang--Mills model in $D$ dimensions:
\begin{lstlisting}[style=pycode]
# models/my_model.py
import torch
from matrix_hmc.models.base import MatrixModel
import matrix_hmc.config as cfg

class MyModel(MatrixModel):
    def __init__(self, dim, ncol, g):
        super().__init__(nmat=dim, ncol=ncol)
        self.g = g
        self.is_hermitian = True
        self.is_traceless = True

    def load_fresh(self):
        self.set_state(torch.zeros((self.nmat, self.ncol, self.ncol),
                                   dtype=cfg.dtype, device=cfg.device))

    def potential(self, X=None):
        X = self._resolve_X(X)
        mass = torch.einsum("bij,bji->", X, X).real
        comm_sq = sum(
            torch.trace((X[i] @ X[j] - X[j] @ X[i])
                        @ (X[i] @ X[j] - X[j] @ X[i])).real
            for i in range(self.nmat) for j in range(i + 1, self.nmat))
        return (self.ncol / self.g) * (mass - 0.5 * comm_sq)

    def measure_observables(self, X=None):
        X = self._resolve_X(X)
        eigs = [torch.linalg.eigvalsh(X[i]).cpu().numpy()
                for i in range(self.nmat)]
        return eigs, None

def build_model(args):
    return MyModel(dim=args.nmat, ncol=args.ncol, g=args.coupling[0])
\end{lstlisting}
For fermionic models, one simply adds the determinant contribution to the potential, and autograd handles the matrix-element derivatives of $\log|\det\mathcal{D}(X)|$ without additional model-specific force code.

The model is then invoked from the command line as
\begin{lstlisting}[style=pycode, language=bash]
matrix-hmc --model ./my_model.py --ncol 20 \
    --coupling 1.0 --nmat 4 --niters 1000 --step-size 0.1 --nsteps 10
\end{lstlisting}
\section{Fit comparison for the Type~I commutativity ratio}
\label{app:type1Fits}

To assess whether the power-law decay $R\sim\lambda^b$ is genuinely the preferred functional form, and not an artefact of the fit range, we fit the ratio $R(\lambda)$ using seven candidate functions. All fits use the same dataset over $\lambda\in[60,450]$; no points are withheld. The candidate functions are

\begin{equation}
\begin{aligned}
f_1(\lambda) = a\,\lambda^b &\quad\qquad f_2(\lambda) = a\,\lambda^b + c &\qquad f_3(\lambda) = a\,(\lambda+d)^b\\
f_4(\lambda) = a\,\lambda^b\!\left(1+\tfrac{c}{\lambda}\right) &\quad\qquad f_5(\lambda) = a\,\lambda^b + c\,\lambda^d &\qquad f_6(\lambda) = \dfrac{a}{1+(\lambda/\lambda_0)^b} \\
& \quad\qquad f_7(\lambda) = a\,e^{-b\lambda}
\end{aligned}
\end{equation}

\begin{figure}[h]
    \centering
    \includegraphics[width=0.82\linewidth]{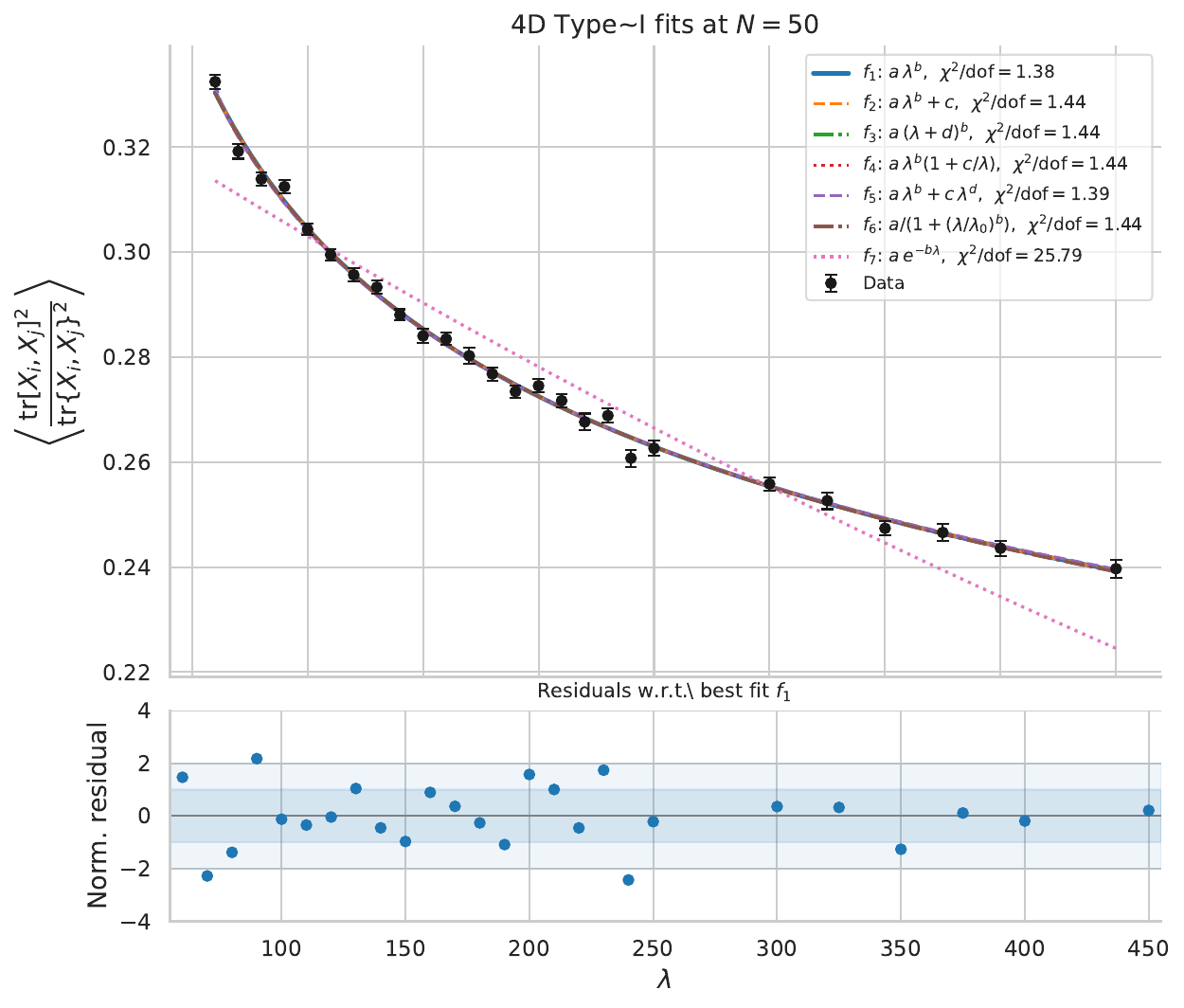}
    \caption{All seven ansatze fitted to the Type~I ratio $R(\lambda)$ at $N=50$. The lower panel shows normalised residuals $(R_i - f_1(\lambda_i))/\sigma_i$ relative to the best-fit power law $f_1$; the bands mark $\pm1\sigma$ and $\pm2\sigma$. }
    \label{fig:type1-allfits}
\end{figure}

Table~\ref{tab:type1fits} collects the best-fit parameters, $\chi^2/\mathrm{dof}$, and $p$-values; the fits are shown graphically in Figure~\ref{fig:type1-allfits}. The main conclusion is that either the other models reduce to the power law, have very large errors in the estimation or have large $\chi^2$/dof. The pure power law $f_1$ achieves the lowest reduced $\chi^2 = 1.38$ ($p=0.10$) among all seven models. The additional parameters in $f_2$--$f_4$ are statistically consistent with zero and do not improve the fit, as reflected in their higher $\chi^2/\mathrm{dof}\approx 1.44$. The two-term form $f_5$ has one very small exponent $b\approx -31.6$, effectively collapsing to a single power law over the data range. The data shows that we do not require any structure beyond $f_1$. The exponential $f_7$ is decisively ruled out at $\chi^2/\mathrm{dof}=25.8$, confirming that the decay is algebraic rather than exponential.

\begin{table}[h]
\centering
\small
\renewcommand{\arraystretch}{1.35}
\begin{tabular}{llp{9.5cm}c}
\toprule
 & Model & Best-fit parameters & $\chi^2/\mathrm{dof}$ \\
\midrule
$f_1$ & $a\,\lambda^b$
  & $a=0.636\pm0.006$,\quad $b=-0.160\pm0.002$
  & \textbf{1.38} \\[2pt]
$f_2$ & $a\,\lambda^b+c$
  & $a=0.636\pm0.030$,\quad $b=-0.160\pm0.039$,\quad $c=(0.0\pm6.8)\times10^{-2}$
  & 1.44 \\[2pt]
$f_3$ & $a\,(\lambda+d)^b$
  & $a=0.643\pm0.026$,\quad $b=-0.162\pm0.007$,\quad $d=1.5\pm5.8$
  & 1.44 \\[2pt]
$f_4$ & $a\,\lambda^b(1+c/\lambda)$
  & $a=0.636\pm0.025$,\quad $b=-0.160\pm0.007$,\quad $c=(0.0\pm9.0)\times10^{-1}$
  & 1.44 \\[2pt]
$f_5$ & $a\,\lambda^b+c\,\lambda^d$
  & $a=(4.0\pm0)\times10^{53}$,\quad $b=-31.6\pm0.1$,\newline $c=0.632\pm0.006$,\quad $d=-0.159\pm0.002$
  & 1.39 \\[2pt]
$f_6$ & $a/(1+(\lambda/\lambda_0)^b)$
  & $a=3.5\pm9.0$,\quad $\lambda_0=(1.3\pm24.6)\times10^{-4}$,\newline $b=0.174\pm0.039$
  & 1.44 \\[2pt]
$f_7$ & $a\,e^{-b\lambda}$
  & $a=0.330\pm0.001$,\quad $b=(8.57\pm0.10)\times10^{-4}$
  & 25.79 \\
\bottomrule
\end{tabular}
\caption{Fit comparison for the Type~I commutativity ratio $R(\lambda)$ at $N=50$, fitted to all 26 data points. The pure power law $f_1$ achieves the lowest $\chi^2/\mathrm{dof}$ with no degeneracies. In $f_2$ and $f_4$ the extra parameter $c$ is consistent with zero, with an uncertainty orders of magnitude larger than the central value. In $f_5$ the first term degenerates ($b\approx-31.6$ makes $a\lambda^b\approx 0$ over the data range, so $a$ is unconstrained), reducing the fit to a single power law. In $f_6$, the $1$ in the denominator can be dropped for the parameter values and we again reduce to a power law; besides, both $a$ and $\lambda_0$ carry errors far exceeding their central values. The exponential $f_7$ is decisively disfavoured.}
\label{tab:type1fits}
\end{table}

\section{The Hoppe model at strong coupling}
\label{app:hoppe}
In this appendix, we review the solution of the Hoppe model at strong coupling and the universality of huge operators in this model, following \cite{Guerrieri:2025ytx}. In this appendix, we will use the notation $X_1=X$ and $X_2=Y$ in terms of which, the action of the Hoppe model is given by (note that we are using the normalization \eqref{eq:bosonicYMAction}, which is different from \cite{Guerrieri:2025ytx}):
\begin{equation}
    S = \frac N\lambda\, \tr \left(X^2 + Y^2 -\frac12[X,Y]^2 \right) \,.
\end{equation}
The model has a single dimensionless coupling $\lambda$. In order to solve the model, we first diagonalize $X$ to obtain the following integral
\begin{equation}
    \int dx\, dY\ \prod_{i<j} (x_i-x_j)^2 \exp\left[-\frac N\lambda\sum_i x_i^2 - \frac N\lambda\sum_{i,j} Y_{ij}Y_{ji} \left(1+\frac{1}2 (x_i-x_j)^2\right) \right] \,.\label{eq:hoppeYprop}
\end{equation}
where the Vandermonde determinant $\prod_{i<j} (x_i-x_j)^2$ arises from the Jacobian of the change of variables as usual, and we have dropped the constant prefactors. We can then integrate out the $Y$ matrices to get
\begin{equation}
    \int dx\ \prod_{i<j} \frac{(x_i-x_j)^2}{2+ (x_i-x_j)^2} \exp\left[-\frac N\lambda\sum_i x_i^2\right] \,.
\end{equation}
The saddle point equation for the eigenvalue density $\rho(x)$ is then given by
\begin{equation}
    x = 2\lambda \fint dy\, \frac{\rho(y)}{(x-y)(2+(x-y)^2)} \,.
\end{equation}
In order to make the singularity structure more transparent, let us rescale the variables as $x = \sqrt{2\lambda} u$ and $y = \sqrt{2\lambda} v$. Then, the saddle point equation becomes
\begin{equation}
    2u = \fint dv\, \frac{\tilde\rho(v)}{(u-v)(1+\lambda(u-v)^2)} \,.
\end{equation}
The integral kernel on the RHS at strong coupling $\lambda \to \infty$ becomes a derivative of the delta function, and we get
\begin{equation}
    2u = -\int dv\, \tilde\rho(v) \frac{\pi}{\sqrt{\lambda}} \delta'(u-v) = -\frac{\pi}{\sqrt{\lambda}} \tilde\rho'(u) \,.
\end{equation}
We can now solve this equation, together with the normalization condition $\int du\, \tilde\rho(u) = 1$, to get the eigenvalue density at strong coupling
\begin{equation}
    \rho(x) = \frac{L_{\texttt{vac}}^2 - x^2}{4L_{\texttt{vac}}^3/3} \,, \quad L_{\texttt{vac}} =  \left(\frac{3\pi\lambda}{\sqrt{2}}\right)^{\frac13} \,.
\end{equation}

We can also compute the expectation value of the commutator squared at strong coupling using the propagator of the $Y$ matrices that follows from \eqref{eq:hoppeYprop},
\begin{align*}
    -\frac1N \langle \tr[X,Y]^2 \rangle 
    &= \frac{\lambda}{N^2}\sum_{i\ne j} \left\langle\frac{(x_i-x_j)^2}{2+(x_i-x_j)^2}\right\rangle = \lambda + \ldots
\end{align*}
We see that the commutator squared is parametrically smaller than $\tr X^4\sim \lambda^{4/3}$. This shows that the Hoppe model is commuting at strong coupling.

Now, let us consider the insertion of a huge operator in the Hoppe model. With the insertion of a classical source $J$ for the matrix $X$, the integral becomes
\begin{equation}
    \int dx\, dU\ \prod_{i<j} \frac{(x_i-x_j)^2}{2+ (x_i-x_j)^2} \exp\left[-\frac N\lambda\sum_i x_i^2 + \frac{N}{\sqrt{\lambda}} \,\tr\, J U X U^\dagger \right]
\end{equation}
where we need to integrate over the unitary group $U(N)$ to account for the angle between the source and $X$. For this, we use the Harish-Chandra-Itzykson-Zuber formula
\begin{equation}
    \int dU\, e^{\beta\, \tr J U X U^\dagger} = \frac{\det_{ij} e^{\beta\, j_i x_j}}{\Delta(j)\Delta(x)} \,,
\end{equation}
For simplicity, let us consider the particular case of a line source studied in section \ref{sec:universality} that is uniformly distributed on a line of size $2\alpha$. For this case, the eigenvalues of the source are given by $j_i = -\alpha + \frac{2i\alpha}N$ and the determinant in the HCIZ formula simplifies greatly to $\det_{ij} e^{\frac{N}{\sqrt{\lambda}}j_i x_j} = e^{-\frac{N}{\sqrt{\lambda}}\tr X}\Delta(e^{\frac{2N}{\sqrt{\lambda}}\alpha x})$. This allows us to solve the saddle point equations once again using the same strong coupling simplification as above. The result is that the eigenvalue density is again given by a parabolic distribution but with a different size of the support $L$:
\begin{equation}
    \rho_{\texttt{huge}}(x) = \frac{L^2 - x^2}{4L^3/3} \,, \quad L =  L_{\texttt{vac}} \times \left(\frac1{1-\frac{\alpha^2}{12}}\right)^{\frac13} \,.
\end{equation}

This shows that, at least for the case of a line source, the eigenvalue density is universal up to a rescaling of the size of the support. As discussed at length in \cite{Guerrieri:2025ytx}, the same story holds for more general sources, and the parabolic shape of the eigenvalue density is universal at strong coupling. Furthermore, we see here that the support diverges at $\alpha=\sqrt{12}$, beyond which we have a non-universal eigenvalue density. 

\section{Integrating out off-diagonals at strong coupling}
\label{app:strong coupling fixed N}
In this appendix, we perturbatively integrate out the off-diagonal modes in a strong-coupling expansion, demonstrate the impact of the fermion number, and use the resulting effective theory for the diagonal modes to compute observables.

\subsection{Method}
We consider a matrix model whose action has the form
\al{
\spl{
S = \frac{N}{\lambda}\mathrm{tr} \Biggl[ -\frac{1}{4} & [X_I,X_J]^2 - \frac{i}{2} \bar{\psi}\Gamma^{I} [X_I,\psi]  + X^{I} X^I + \bar \psi \psi \Biggr] \,. \label{eq: generic YM action dup}
}
}
where $\psi$ is a spinor transforming covariantly under rotations. This notation is adapted to Majorana spinors where $\bar \psi = \psi^\top C$ is related to $\psi$. For Dirac spinors, we typically use $-i \bar \psi \Gamma^I [X_I, \psi]$ where $\bar \psi$ and $\psi$ are treated as independent. However, the precise factor of 2 will not matter for the discussion and the observables we will consider. Moreover, note that $\psi$ can also encompass a flavor index. Namely, $\psi$ can be thought of as a vector of $\mathsf f$ Majorana or Dirac spinors, and we define $\mathcal{N}$ as the total number of real fermionic components. Having $\mathsf f$ fermions means that the Pfaffian is raised to an integer power $\mathsf f$. As mentioned in the main text, we can now allow $\mathsf f$ to be non-integer, meaning $\mathcal{N}$ can be non-integer. Thus, all the results we will derive in this appendix only depend on the value of $D \in \mathbb{N}$ and $\mathcal{N} \geq 0$.

As explained in section \ref{sec: effective theory}, we split the matrices into diagonal and off-diagonal parts,
\begin{equation}
    X_I = r_I + q_I \,, \qquad \psi_\alpha = \theta_\alpha + \Theta_\alpha
\end{equation}
where $r_I, \theta_\alpha$ are diagonal matrices and $q_I, \Theta_\alpha$ are off-diagonal matrices. We further define
\begin{equation}
    r_{ab}^I\equiv r_a^I-r_b^I \,, \quad \hat{r}_{a b}^I \equiv r_{a b}^I/|r_{a b}| \,, \quad |r_{a b}| \equiv \sqrt{\sum_I r_{a b}^I r_{a b}^I} \,,
\end{equation}
where $r_a^I$ ($a=1,\ldots,N$) are the components of the diagonal matrix $r^I$. 

\paragraph{Gauge fixing procedure:} Using $SU(N)/U(1)^{N-1}$ transformations, we fix $(N^2-N)$ degrees of freedom by imposing the gauge fixing constraint
\begin{equation}
    F_{ab} \equiv\sum_I r^I_{ab} q_{ab}^I \overset{!}{=} 0 \qquad (a \neq b)
\end{equation}
As is standard in gauge-fixing gauge theories, this is done by inserting in the path integral 
\begin{equation}
    \delta(F_{ab}) \Biggl| \det \left( \frac{\delta F_{ab}}{\delta \alpha_{cd}} \right)\Biggl|_{\alpha=0}
\end{equation}
up to a gauge group volume factor, where the variation is done under an infinitesimal gauge transformation $X^I \to X^I + i [\alpha, X^I]$. Under such a transformation, the diagonal and off-diagonal components transform as
\begin{equation}
    \delta r_a^I = i \alpha_{a b} q_{b a}^I - i q_{a b}^I \alpha_{b a}
\end{equation}
\begin{equation}
    \delta q_{a b}^I = - i \alpha_{a b} r_{a b}^I + i \alpha_{a c} q_{c b}^I - i q_{a c}^I \alpha_{c b}
\end{equation}
Using this, it is straightforward to obtain the derivative of $F$ which gives
\begin{equation}
    \frac{\delta F_{ab}}{\delta \alpha_{cd}} = -i |r_{ab}|^2 (\delta_{ac} \delta_{bd} + K_{ab,cd})
\end{equation}
where
\begin{equation}
    K_{ab;cd} \equiv \frac{r_{a b}^I}{|r_{a b}|^2} (q_{a c}^I \delta_{b d} - \delta_{a c} q_{d b}^I) + \frac{q_{a b}^I q_{d c}^I}{|r_{a b}|^2} (\delta_{a d} - \delta_{a c} - \delta_{b d} + \delta_{b c}) \label{eq: K definition}
\end{equation}
The factor to insert in the partition function is thus
\begin{align}
\begin{split}
    \delta(F_{ab}) \Biggl| \det \left( \frac{\delta F_{ab}}{\delta \alpha_{cd}} \right)\Biggl|_{\alpha=0} & = \delta(r_{ab}^I q_{ab}^I)  |\det(|r_{ab}|^2 (\mathbb{1} + K))| \\ & = \delta(\hat r_{ab}^I q_{ab}^I) \left(\prod_{a \neq b} |r_{ab}| \right)|\det(\mathbb{1} + K)|
\end{split}
\end{align}
\paragraph{Expansion of the action:} Thus, the full partition function is written as
\begin{equation}
    Z \propto \int [dr] [dq] [d \theta] [d \Theta] \delta\left( \sum_a r_a^I \right) \delta(\hat r_{ab}^I q_{ab}^I) \left(\prod_{a \neq b} |r_{ab}|\right) |\det (\mathbb 1 + K)| e^{-S} \label{eq: part func in SO(D) preserving gauge}
\end{equation}
where $K$ is a $(N^2-N) \times (N^2-N)$ matrix defined in \eqref{eq: K definition}.
Expanding $S$ in diagonal and off-diagonal modes, we obtain in the gauge $F_{ab} = 0$,
\begin{equation}
    S = S_\mathrm{quad} + S_\mathrm{int}  \label{eq: full off-diag expansion}
\end{equation}
\begin{align}
\begin{split}
    S_\mathrm{quad} & = \frac{N}{\lambda} |r_a|^2 + \frac{N}{\lambda} \bar \theta_a \theta_a  +\frac{N}{\lambda} \left(- \frac{i}{2} r_{a b}^I \bar\Theta_{a b} \Gamma^I \Theta_{b a} + \bar \Theta_{a b}\Theta_{b a} \right) \\ & \quad + q_{ab}^I \left( \frac{N}{2\lambda}|r_{ab}|^2 \delta^{IJ} + \frac{N}{\lambda} \delta^{IJ} \right) q_{ba}^J 
\end{split}
\end{align}
\begin{align}
\begin{split}
S_\mathrm{int} & = \frac{N}{\lambda} \Bigl\{- 2 r_{a b}^I q_{a b}^J q_{b c}^I q_{c a}^J - \frac{1}{2} (q_{a b}^I q_{b c}^J q_{c d}^I q_{d a}^J - q_{a b}^I q_{b c}^J q_{c d}^J q_{d a}^I)  \\
& \qquad +i \bar \Theta_{a b} \Gamma^I q_{b a}^I (\theta_a - \theta_b) -i q_{a b}^I \bar \Theta_{b c} \Gamma^I \Theta_{c a} \Bigr\} 
\end{split}
\end{align} 

\paragraph{The large coupling limit:}
As mentioned in the main text, the large coupling limit is best understood by rescaling the variables to
\begin{equation}
    r_I \equiv \sqrt{\lambda} x_I, \qquad \theta_\alpha \equiv \sqrt{\lambda} \eta_\alpha \,, \qquad \Theta_\alpha = \lambda^{1/4} \chi_\alpha \,.
\end{equation}
In these coordinates, we have
\begin{equation}
    K \to 0, \qquad S_\mathrm{int} \to 0 \qquad(\lambda \to \infty, x \ \text{fixed})
\end{equation}
meaning that the partition function reduces to
\begin{align}
\begin{split}
    Z & \propto \int [dx] [dq][d \eta] [d \chi] \delta\left( \sum_a x_a^I \right) \delta(\hat x_{ab}^I q_{ab}^I) \left(\prod_{a \neq b} |x_{ab}|\right) \\ & \qquad \exp\left(-N \Bigl[x_a^I x_a^I + \bar \eta_a^\alpha \eta_a^\alpha + \frac{1}{2}  |x_{ab}|^2 |q_{ab}|^2 - \frac{i}{2} x_{a b}^I \bar \chi_{ba} \Gamma^I \chi_{ab} + \mathcal{O}(\lambda^{-1/4}) \Bigr]\right)
\end{split}
\end{align}
Ignoring $\mathcal{O}(\lambda^{-1/4})$, we simply integrate out $q$'s and $\chi$'s using
\begin{equation}
    \int [dq^I] \delta(\hat x_{a b} \cdot q_{a b}) e^{- \frac{N}{2} |x_{a b}|^2 q_{a b}^I q_{b a}^I} \propto \prod_{a \neq b} |x_{a b}|^{- (D-1)} \,,
\end{equation}
\begin{equation}
    \int [d \chi_\alpha] e^{\frac{iN}{2} x_{a b}^I \bar \chi_{b a} \Gamma^I \chi_{a b}} \propto\prod_{a \neq b} |x_{a b}|^{\mathcal{N}/2} \,.
\end{equation}
Combining with the Vandermonde term, and going back to the original $r,\theta$ coordinates, we obtain
\begin{equation}
    Z \propto \int\prod_{a=1}^N \left( \prod_{I=1}^D dr_I^a \prod_{\alpha=1}^{\mathcal{N}}d\theta_\alpha^a \right) \delta^{(D)}\left(\sum_a r_a \right) \prod_{a <b} |r_{ab}|^{\mathcal{N}-2(D-2)} \mathrm{exp}(-S_\mathrm{diag}) \,,
\end{equation}
where $S_\mathrm{diag}$ is the original action restricted to diagonal matrices,
\begin{equation}
     S_\mathrm{diag} =\frac{N}{\lambda}( r_I^a r_I^a + \bar \theta^a_\alpha   \theta^a_\alpha) \,.
\end{equation}

To summarize, we obtain that for a gauge invariant function $f$,
\begin{equation}
    \langle f(X_I) \rangle \xrightarrow{\lambda \to \infty} \langle f(r_I) \rangle_\mathrm{diag}  \,,\label{eq: observables massless limit 2}
\end{equation}
where
\begin{equation}
    \langle f(r_I) \rangle_\mathrm{diag} \equiv \frac{\int\prod_{a=1}^N \left( \prod_{I=1}^D dr_I^a  \right) f(r_I) \delta^{(D)}\left(\sum_a r_a \right) \prod_{a <b} |r_{ab}|^{\mathcal{N}-2(D-2)} \mathrm{exp}(-S_\mathrm{diag})}{\int\prod_{a=1}^N \left( \prod_{I=1}^D dr_I^a  \right) \delta^{(D)}\left(\sum_a r_a \right) \prod_{a <b} |r_{ab}|^{\mathcal{N}-2(D-2)} \mathrm{exp}(-S_\mathrm{diag})} \label{eq: diag eff action}
\end{equation}

\paragraph{Some observables:} Finally, let us use these methods to compute a few observables in this fixed $N$ strong coupling regime. We can compute moments of a given matrix $X_1$, such as
\begin{equation}
    \frac{1}{N} \langle \mathrm{tr} X_1^2 \rangle_{SU(N)} \xrightarrow{\lambda \to \infty} \lambda \left(\frac{N-1}{2N^2} +\frac{\mathcal{N}-2(D-2)}{4D} \frac{N(N-1)}{N^2}\right) \label{eq: moment at large lambda finite N}
\end{equation}
The commutator can be obtained integrating out the off-diagonal modes in their Gaussian approximation, yielding
\begin{equation}
    -\frac{1}{N} \langle \mathrm{tr} [X_1, X_2]^2 \rangle \xrightarrow{\lambda \to \infty} \frac{2\lambda}{D} \frac{N(N-1)}{N^2} \label{eq: CommSquaredApprox}
\end{equation}
Finally, the commutator-anticommutator ratio can also be obtained in the large $\lambda$ limit. One obtains
\begin{equation}
    \left\langle \frac{-\sum_{{I< J}}\mathrm{tr} [X_I,X_J]^2}{\sum_{{I< J}}\mathrm{tr}\{ X_I, X_J \}^2} \right\rangle_{SU(N)} \xrightarrow{\lambda \to \infty} \frac{D-1}{4 \lambda}N^2 (N-1)\left\langle \frac{1}{\sum_{{I< J}}\sum_a (r_a^I)^2(r_a^J)^2} \right\rangle_0
\end{equation}
where
\al{
\spl{
\Biggl\langle & \frac{1}{\sum_{{I< J}}\sum_a (r_a^I)^2(r_a^J)^2} \Biggr\rangle_0 \\ & \qquad = \frac{\int\prod_{a=1}^N d^D r_a (\sum_{{I< J}}\sum_a (r_a^I)^2(r_a^J)^2)^{-1} \delta^D(\sum_a \vec r_a) (\prod_{a \neq b} |r_{ab}|^\nu)e^{- \sum_a \vec r_a^2}}{\int \prod_{a=1}^N d^D r_a \delta^D(\sum_a \vec r_a) (\prod_{a \neq b} |r_{ab}|^\nu)e^{- \sum_a \vec r_a^2}}
}
}
\begin{equation}
    \nu \equiv \frac{\mathcal N}{2} - (D-2)
\end{equation}
In particular for the cases of interest, we can evaluate this integral numerically at finite $N$ with the result
\begin{equation}
    \left\langle \frac{-\sum_{I<J} \tr [X_I,X_J]^2}{\sum_{I<J} \tr \{X_I,X_J\}^2} \right\rangle \xrightarrow{\lambda \to \infty} \begin{cases}
        2.52 \lambda^{-1} & D=4, \mathcal{N} =8, N=30 \ (4d \ \text{adj QCD}) \\ 9.6 \lambda^{-1} & D=5, \mathcal{N}=8, N=30 \ (5d \ \text{adj QCD}) \\ 1207 \lambda^{-1} & D=3, \mathcal{N}=2, N=40 \ (3d \ \mathrm{SYM}) \\ 1344 \lambda^{-1} & D=4, \mathcal{N}=4, N=50 \ (4d \ \text{type I})
    \end{cases}
\end{equation}
Note that the last two entries correspond to the critical case $\mathcal{N} = \mathcal{N}_c$, for which $R \sim \frac{\lambda}{\lambda^2/N^2} \sim N^2 \lambda^{-1}$, which results in the huge prefactors. 

\subsection{Joint densities for the supercritical models}
As mentioned above, and as supported by the numerical results, the infinite $N$ limit of supercritical matrix models that have $\mathcal{N} > 2(D-2)$ fermions can be described by the diagonal effective theory when $\lambda \gg 1$. Here, we solve the $N \to \infty$, $\lambda \gg 1$ limit of the supercritical models by finding the large $N$ densities that are saddles of \eqref{eq: diag eff action}. 

In the original coordinates \eqref{eq: generic YM action dup}, the large $N$ saddle minimizes
\begin{equation}
    I[\rho]=\frac1\lambda\int d^D r\,\rho(r)|r|^2
    -\nu\int d^D r\,d^D s\,\rho(r)\rho(s)\log|r-s|\,.
\end{equation}
The corresponding saddle-point equation is
\begin{equation}
    \frac{|r|^2}{\lambda}
    -2\nu\int d^D s\,\rho(s)\log|r-s|=\ell\,,
    \qquad r\in \operatorname{supp}\rho\,,
    \label{eq:diag-density-saddle}
\end{equation}
where $\ell$ is a Lagrange multiplier constant. By rotational invariance, the saddle is radial.

For $D=2$ and $D=3$, the positive solution has support on a ball. It is
\begin{equation}
    \rho_{D,\nu}(r)
    =
    \frac{\left(R_D^2-|r|^2\right)^{1-\frac D2}}
    {\pi^{D/2}\Gamma\left(2-\frac D2\right)R_D^2}\,
    \Theta(R_D-|r|)\,,
    \qquad
    R_D^2=\frac{2\nu\lambda}{D}\,.
    \label{eq:diag-density-bulk}
\end{equation}
Indeed, this density is normalized and equation \eqref{eq:diag-density-saddle} becomes
\begin{equation}
    \int d^D s\,\rho_{D,\nu}(s)\log|r-s|
    =C_{D,\nu}+\frac{|r|^2}{D R_D^2}\,,
    \qquad |r|<R_D\,,
\end{equation}
with some constant $C_{D,\nu}$ so that \eqref{eq:diag-density-saddle} fixes precisely the radius in \eqref{eq:diag-density-bulk}. 

For $D\geq4$, the large $N$ density is the uniform measure on a sphere,
\begin{equation}
    \rho_{D,\nu}(r)
    =
    \frac{\delta(|r|-R_D)}{\Omega_{D-1}R_D^{D-1}}\,,
    \qquad
    R_D^2=\frac{\nu\lambda}{2}\,,
    \qquad
    \Omega_{D-1}=\frac{2\pi^{D/2}}{\Gamma(D/2)}\,.
    \label{eq:diag-density-shell}
\end{equation}
The radius $R_D$ is fixed by differentiating the saddle-point equation with respect to the position \(r\):
\begin{equation}
    \frac{2r}{\lambda}
    -2\nu\int d^Ds\,\rho(s)\frac{r-s}{|r-s|^2}=0 .
\end{equation}
For a uniform shell \(r=R_D\hat n\), \(s=R_D\hat m\), rotational symmetry implies that the integral is radial. Its radial projection is
\begin{equation}
    \hat n\cdot \frac{r-s}{|r-s|^2}
    =
    \frac{1}{2R_D},
\end{equation}
and hence
\begin{equation}
    \int d^Ds\,\rho(s)\frac{r-s}{|r-s|^2}
    =
    \frac{r}{2R_D^2}.
\end{equation}
This implies
\begin{equation}
    \frac{2R_D}{\lambda}-\frac{\nu}{R_D}=0 .
\end{equation}
As a side-remark, note that the solution \eqref{eq:diag-density-shell} is a saddle-point in any $D$, in particular, also in the case $D=2,3$. However, in these cases, this solution is unstable, and the on-shell action is minimized by \eqref{eq:diag-density-bulk}. To see this unstability explicitly, define
\begin{equation}
    \Phi_\rho[r] = \frac{|r|^2}{\lambda} -2 \nu \int d^D s \rho(s) \log |r-s|
\end{equation}
The saddle-point equation \eqref{eq:diag-density-saddle} is $\Phi_\rho[r] = \ell$ when $r$ is in the support of $\rho$, namely $r \in \mathrm{supp} \ \rho$. However, for $\rho$ to be an actual minimum, one needs to verify the condition $\Phi_\rho[r] \geq \ell$ when $r \notin \mathrm{supp} \ \rho$. To understand this, consider
\begin{equation}
    \rho_\epsilon(r) \equiv (1- \epsilon) \rho(r) + \epsilon \delta^{(D)}(r-r_0)
\end{equation}
where $r_0 \notin \mathrm{supp} \ \rho $. Then, at leading order in $\epsilon$ and using $\Phi_\rho[r] = \ell$ in the support of $\rho$, we find
\begin{equation}
    I[\rho_\epsilon]- I[\rho] = \epsilon (\Phi_\rho [r_0] - \ell)
\end{equation}
When this is negative, $\rho$ is unstable and prefers emitting eigenvalues. This is what happens in $D=2,3$, and the true minimum configuration is the ball-supported solution rather than the spherical-shell solution.

Given the large $N$ density $\rho_{D,\nu}$, any single-trace bosonic observable has a simple large $N$ limit, namely
\begin{equation}
    \frac{1}{N} \langle \mathrm{tr} f(X_I) \rangle \xrightarrow[N\to\infty]{\lambda\to\infty}\int d^D r\,\rho_{D,\nu}(\vec r)f(r_I) \,,
\end{equation}
where the ordering of the matrices $X_I$ inside $f$ is irrelevant since matrices commute in this limit.

For example, in all dimensions $D\geq2$ and for all $\nu>0$,
\begin{equation}
    \frac1N\langle \tr X_I^2\rangle
    \xrightarrow[N\to\infty]{\lambda\to\infty}
    \frac{\nu\lambda}{2D}
    =
    \frac{\mathcal N-\mathcal N_c}{4D}\lambda\,,
\end{equation}
in agreement with the finite $N$ result \eqref{eq: moment at large lambda finite N}.

The same rule also gives the leading anticommutator contribution to the commutativity ratio. We get
\begin{equation}
    \frac1N
    \left\langle \sum_{I<J}\tr\{X_I,X_J\}^2\right\rangle
    =
    \begin{cases}
    \displaystyle
    \frac{\nu^2\lambda^2(D-1)}{3D}\,, & D=2,3\,,\\[6pt]
    \displaystyle
    \frac{\nu^2\lambda^2(D-1)}{2(D+2)}\,, & D\geq4\,.
    \end{cases}
\end{equation}
Combining this with \eqref{eq: CommSquaredApprox}, one obtains for $D\geq4$
\begin{equation}
    \left\langle\frac{
    -\sum_{I<J}\tr[X_I,X_J]^2
    }{
    \sum_{I<J}\tr\{X_I,X_J\}^2
    }
    \right\rangle\xrightarrow[N\to\infty]{\lambda\to\infty}
    \frac{2(D+2)}{\nu^2}\,\lambda^{-1}\,.
\end{equation}
Thus the $D=4$ and $D=5$ adjoint QCD models, for which $(D,\nu)=(4,2)$ and $(D,\nu)=(5,1)$ respectively, give
\begin{equation}
    \left\langle\frac{
    -\sum_{I<J}\tr[X_I,X_J]^2
    }{
    \sum_{I<J}\tr\{X_I,X_J\}^2
    } \right\rangle
    \xrightarrow[N\to\infty]{\lambda\to\infty}
    \begin{cases}
        3\,\lambda^{-1}\,, & D=4,\ \mathcal N=8\,,\\
        14\,\lambda^{-1}\,, & D=5,\ \mathcal N=8\,.
    \end{cases}
\end{equation}

\section{Non-universality in supercritical models}
\label{app: non-universal supercritical}
We have seen that supercritical models admit a diagonal description in the strong 't Hooft coupling limit. In this regime the eigenvalue cloud has width \(O(\sqrt{\lambda})\). A huge deformation of the form
\[
    O_{\rm huge}\sim
    \exp\!\left[N\,\tr V(X/\sqrt{\lambda})\right]
\]
therefore contributes \(O(N^2)\) to the diagonal saddle-point action. For example,
\[
    N\,\tr \left(\frac{X}{\sqrt{\lambda}}\right)^{2k}
    \sim
    \frac{N}{\lambda^k}\tr X^{2k}
    \sim O(N^2),
\]
which is of the same order as the original mass term. Thus, unlike in the critical models, the detailed form of \(V\) is expected to affect the limiting joint density. In this appendix we demonstrate this non-universality in two solvable examples.

\subsection{Quartic deformation}
Start from a supercritical theory \eqref{eq: generic YM action dup} with $\mathcal N > \mathcal N_c$ deformed by a quartic bosonic term $ \sim g NX^4 / \lambda^2$ where $g$ is a deformation parameter, namely,
\al{
\spl{
S_\mathrm{quartic} = \frac{N}{\lambda}\mathrm{tr} \Biggl[ -\frac{1}{4} & [X_I,X_J]^2 - \frac{i}{2} \bar{\psi}\Gamma^{I} [X_I,\psi]  + X^{I} X^I + \frac{g}{\lambda} X_I X_I X_J X_J + \bar \psi \psi \Biggr] \,.
}
}
Then, the method of integrating out the off-diagonals of appendix \ref{app:strong coupling fixed N} can be applied, and the large $N$ and large $\lambda$ limit can be described by continuous joint densities that minimize the effective action
\begin{equation}
    S_\mathrm{eff}=\frac1\lambda\int d^D r\,\rho(r) \left(r^2 + \frac{g r^4}{\lambda}\right)
    -\nu\int d^D r\,d^D s\,\rho(r)\rho(s)\log|r-s|\,.
\end{equation}
where $\nu = (\mathcal{N}-\mathcal{N}_c)/2 >0$. Before solving the saddle-point, it is illuminating to rescale out $\lambda$ by $r = \sqrt{\lambda} x$ and $\rho(r) = \frac{1}{\lambda^{D/2}} f(x)$, yielding (up to an irrelevant constant)
\begin{equation}
    S_\mathrm{eff} = \int d^D x f(x) V(x) - \nu \int d^D x d^D y f(x) f(y) \log |x-y| \label{eq: effective potential deformed diagonal theory}
\end{equation}
where now $V(x)$ has no $\lambda$-dependence,
\begin{equation}
    V(x) = x^2 + g x^4
\end{equation}
\paragraph{When $2\leq D < 4$:} There are two regimes, $g \leq g_c$ and $g \geq g_c$ where
\begin{equation}
    g_c = \frac{D^2 (4-D)}{4 \nu (D+2) (D-2)^2}
\end{equation}
When $g \leq g_c$, the solution can be found analytically with the result
\begin{equation}
    f_D(x)=
    \frac{(L^2-x^2)^{1-\frac D2}}
    {\nu\,\pi^{D/2}\Gamma\!\left(2-\frac D2\right)}
    \left[
    \frac D2
    -
    \frac{D(D+2)(D-2)}{2(4-D)}\,gL^2
    +
    \frac{D(D+2)}{4-D}\,g x^2
    \right]
    \Theta(L - |x|).
\end{equation}
where
\begin{equation}
    L^2=
    \frac{
    \sqrt{1+\frac{4(D+2)}{D}g\nu}-1
    }{
    (D+2)g
    }.
\end{equation}
For any fixed nonzero \(g\), the functional form of the density is modified, rather than being related to the \(g=0\) density by a simple rescaling. This provides a first explicit example of non-universality in supercritical models.
When $g \geq g_c$, the solution is supported on a shell $0<x_- < x < x_+$ and cannot be analytically expressed fully explicitly.
\paragraph{When $D \geq 4$:} In this case, turning on $g\neq 0$ does not change the qualitative form of the equilibrium distribution. The density remains supported on a spherical shell, only its radius is shifted. Explicitly,
\begin{equation}
    f_D(\vec x)
    =
    \frac{\delta(|\vec x|-L)}
    {\Omega_{D-1}L^{D-1}},
    \qquad D>4,
\end{equation}
with 
\begin{equation}
    L^2
    =
    \frac{\sqrt{1+4g\nu}-1}{4g}.
\end{equation}
Thus these would remain ``universal'' under $SO(D)$-preserving quartic deformations. We now show that they are not in the case of anisotropic deformations in the next subsection.
\subsection{Anisotropic deformation}
Consider again the effective theory \eqref{eq: effective potential deformed diagonal theory} but in the case of an anisotropic deformation
\begin{equation}
    V(x) = x^2 + g Y_2(x)
\end{equation}
where $Y_2$ is a rank-2 spherical harmonic
\begin{equation}
    Y_2(x) = \frac{x_1^2 - x_2^2}{|x|^2} \label{eq: sph harmonic deformation}
\end{equation}

When $D > 4$, the $g=0$ solution is supported on a spherical shell. Since $\nabla Y_2$ has no radial component, one expects the deformation to affect the angular distribution, but not affect the support of the density which should remain a spherical shell. Indeed, the solution for finite but small enough $g$ is expressed as
\begin{equation}
    f_D(\vec x)
    =
    \frac{\delta(|\vec x|-L)}
    {\Omega_{D-1}L^{D-1}}
    \left[
        1-\frac{D(D-1)g}{\nu}
        \left(
            \frac{x_1^2-x_2^2}{|\vec x|^2}
        \right)
    \right],
    \qquad D>4 .
\end{equation}
where $L$ is still $g$-independent,
\begin{equation}
    L^2=\frac{\nu}{2},
\end{equation}
Note however that the angular distribution is clearly non-universal.

The restriction \(D>4\) is important. The anisotropic deformation \eqref{eq: sph harmonic deformation} is harmless on a shell, because the support stays away from the origin. This is
why the \(D>4\) solution can still be described by a deformed angular density on
a fixed spherical shell. By contrast, for \(2\leq D<4\), the undeformed
equilibrium measure has support reaching the origin, where \(Y_2\) is not smooth.
Equivalently, although \(Y_2\) is bounded, its Laplacian is singular,
\begin{equation}
    \Delta Y_2(x) \sim \frac{Y_2(x)}{|x|^2}.
\end{equation}
Thus a perturbative bulk solution on the undeformed support would require a
singular anisotropic correction to the density, formally behaving as
\begin{equation}
    f_1(x)\sim \frac{Y_2(x)}{|x|^D},
\end{equation}
which is not locally integrable at the origin. Therefore the simple analytic
ansatz used above breaks down for \(2\leq D<4\). We will not analyze this more
complicated problem here.

\bibliographystyle{utphys} 
\bibliography{refs}

\end{document}